\documentclass{aa}  

\usepackage{graphicx}
\usepackage{placeins}
\usepackage{sidecap}
\usepackage{txfonts}
\begin{document}

   \title{Revised Gaia Data Release 2 passbands\thanks{Table~2 is only available at the CDS via anonymous ftp to {\tt cdsarc.u-strasbg.fr~(130.79.128.5)} or via {\tt http://cdsarc.u-strasbg.fr/viz-bin/qcat?J/A+A/vol/page}.}}

   \author{M. Weiler
             }

   \institute{Departament de F{\'i}sica Qu{\`a}ntica i Astrof{\'i}sica, Institut de Ci{\`e}ncies del Cosmos (ICCUB), Universitat de Barcelona (IEEC-UB), Mart{\'i} i Franqu{\`e}s 1, E 08028 Barcelona, Spain\\
              \email{mweiler@fqa.ub.edu}
             }

   \date{Received 21 May 2018; accepted 14 June 2018}

 
  \abstract
   {The European Space Agency mission {\it Gaia} has published, with its second data release (DR2), a catalogue of photometric measurements for more than 1.3 billion astronomical objects in three passbands. The precision of the measurements in these passbands, denoted {\it G}, $G_{\rm BP}$, and $G_{\rm RP}$, reach down to the milli-magnitude level. The scientific exploitation of this data set requires precise knowledge of the response curves of the three passbands.} 
   {This work aims to improve the exploitation of the photometric data by deriving an improved set of response curves for the three passbands, allowing for an accurate computation of synthetic {\it Gaia} photometry.}
   {This is achieved by formulating the problem of passband determination in a functional analytic formalism, and linking the photometric measurements with four observational, one empirical, and one theoretical spectral library.}
   {We present response curves for {\it G}, $G_{\rm BP}$, and $G_ {\rm RP}$ that differ from the previously published curves, and which provide a better agreement between synthetic {\it Gaia} photometry and {\it Gaia} observations.}
{}
 
   \keywords{astronomical databases -- catalogues --  instrumentation: photometers -- techniques: photometric -- techniques: spectroscopic
               }

   \maketitle
%

\section{Introduction}
The European Space Agency (ESA) space mission {\it Gaia} \citep{Gaia2016a} performs an all-sky astrometric, photometric, and spectroscopic survey. Its second data release ({\it Gaia} DR2 in the following, \citet{Gaia2018a}) includes photometric measurements in three different passbands. {\it Gaia's} {\it G} passband covers a wavelength range from the near ultraviolet (roughly 330 nm) to the near infrared (roughly 1050 nm). The other two passbands, denoted $G_{BP}$ and $G_{RP}$, cover smaller wavelength ranges, from approximately 330 to 680 nm, and 630 to 1050 nm, respectively. {\it Gaia} DR2 includes photometric observations of some 1.7 billion astronomical objects in {\it G} band, and more than 1.3 billion objects in $G_{\rm BP}$ and $G_{\rm RP}$ in a magnitude range from about three to 21. For the sources with a $G$ magnitude below 15, the precision of the photometric data reaches the level of some milli-magnitudes \citep{Evans2018}.\par
Interpreting this unique rich data set requires knowledge of the response curves of the passbands in which the photometric observations were performed. Before the launch of the {\it Gaia} spacecraft in 2013, expected response curves based on laboratory measurements and simulations of the instrumental components (mirror reflectance, prism transmissivities, charge-coupled device (CCD) quantum efficiencies) were published by \cite{Jordi2010}. Differences between the predicted passbands and the actual {\it Gaia} DR2 passbands have, however, been detected, and \cite{Evans2018} present updated passbands for {\it G}, $G_{\rm BP}$, and $G_{\rm RP}$ that allow for a more accurate reproduction of the observed photometry by synthetic photometry from spectral energy distributions (SEDs) of astronomical calibration sources. In fact two sets of response curves are provided by \cite{Evans2018}, one set named {\it DR2}, which was used in the preparation of the {\it Gaia} DR2, and one named {\it REV}, which was derived using more accurate knowledge of the instrument, and which is therefore considered to be closer to the true passband than the {\it DR2} set of response curves. In this work, we aim to further improve the {\it Gaia} DR2 passbands. The improvements are achieved in two ways. First, we make use of the techniques for passband reconstruction developed by \cite{Weiler2018}, allowing for a deeper insight into the problem and more control over the passband solutions that can be obtained. Second, we include systematic effects in the {\it Gaia} DR2 photometry described by \cite{Evans2018} and \cite{Arenou2018} in the problem of passband determination.\par
The response curves presented by \cite{Evans2018} have been derived by first generating an initial guess based on measured reflectivity, transmissivity, and quantum efficiency curves plus an adjustment for the position of the wavelength cut-on/off position for $G_{\rm RP}$ and $G_{\rm BP}$. This initial guess has then been modified either by adding a linear combination of the leading few principal components derived from a set of random generated response curves (for {\it G} and $G_{\rm BP}$), or by multiplying the initial guess with a polynomial (for $G_{\rm RP}$) until a good agreement between the synthetic photometry of a set of calibration sources and their observed photometry was achieved (see Chapter 5.3.6 in the {\it Gaia} DR2 online documentation\footnote{\url{http://gaia.esac.esa.int/documentation/}.}). In this work we use a functional analytic formulation of the problem as outlined by \cite{Weiler2018}. As an additional constraint on the shape of the response curves, we make use of empirical stellar spectra and compare their positions in colour-colour diagrams with the distribution of {\it Gaia} DR2 sources in the same colour-colour diagrams. Employing these techniques, we are able to improve the passbands for {\it Gaia} DR2, allowing for a better description of the {\it Gaia} photometric system.\par
As during this work have to refer to the three {\it Gaia} passbands, and for each of the passbands we have to refer to three different solutions for the response curves (the {\it DR2}, the {\it REV,} and the solution from this work), and for each of these cases we have to discriminate between observational magnitudes and magnitudes from synthetic photometry, we simplify the nomenclature in the following way. Instead of referring to $G_{\rm BP}$ and $G_{\rm RP}$, we simply use $BP$ and $RP$ when referring to passbands in a generic way. In the superscript we use {\it obs} and {\it com} to refer to observational magnitudes and computed magnitudes. In the subscript we use {\it dr2}, {\it REV}, and {\it c} to refer to the two passbands by \cite{Evans2018} and the solution from this work, respectively. For the observational photometry, the difference between {\it dr2}, {\it REV,} and {\it c} arises only from small differences in the photometric zero points, except for $G_c$, as discussed in Sect.~\ref{sec:G}.\par
In Sect. \ref{sec:theory} of this work we summarise the mathematical approach in brief and discuss modifications with respect to the work by \cite{Weiler2018}. In Sect. \ref{sec:data} we discuss the calibration data used in detail, and in Sects. \ref{sec:G} to \ref{sec:RP} we derive new passbands for {\it G}, $BP$, and $RP$, compare our results with the initially expected passbands by \cite{Jordi2010} and the revised passbands by \cite{Evans2018}, and compare colour-colour relations predicted with different passbands with observed colour-colour relations. We close this work with a summary and discussion in Sect. \ref{sec:summary}.

\section{Theoretical approach \label{sec:theory}}

\subsection{Mathematical formalism \label{sec:math}}

Following the approach outlined in \cite{Weiler2018}, we can formulate the problem of finding a passband given a set of astronomical objects with well known SED and the photometric observations of these objects as follows. We assume that we have a photon-counting device, such as a CCD detector, that records photo-electrons. Let $p(\lambda)$ be the response curve, defined as the ratio of recorded photo-electrons over the number of photons entering into the instrument, as a function of wavelength $\lambda$. We assume that $p(\lambda)$ is different from zero only inside an a-priori known wavelength interval $I=[\lambda_0,\lambda_1]$. Let $\bf c$ be a vector of length $N$ containing the numbers of observed photo-electrons for $N$ different astronomical objects. The entries of $\bf c$, $c_i,\; i=1,\ldots,N$ are thus the numbers of electrons per unit of time and unit of area. For {\it Gaia} DR2, the unit of area is the aperture of the {\it Gaia} telescope, which is 0.7278~m$^{\rm 2}$. Furthermore we assume that we know the SED of the $N$ objects, to which we refer as the ''calibration sources'', on the wavelength interval $I$. We transform the SED into the spectral photon distribution (SPD) by dividing the SED with the energy of a photon as a function of wavelength. The SPDs of the $N$ calibration sources we denote as $s_i(\lambda) , \; i=1,\ldots,N$. With the abbreviation $\langle\,f \, | \, g \, \rangle := \int_{\lambda_0}^{\lambda_1} f(\lambda)\cdot g(\lambda)\, {\rm d}\lambda$ , the relationship between $c_i$ and $s_i(\lambda)$ is
\begin{equation}
c_i = \langle\, p\, | \, s_i \, \rangle \quad . \label{eq:1}
\end{equation}
Assuming square-integrability for all relevant functions over the wavelength interval $I$, we exploit the properties of a Hilbert space of square-integrable functions over $I$ and the field of real numbers, ${\mathcal L}^2(I)$. The expression $\langle \,f  \, | \, g \, \rangle$ in this context is the scalar product between two vectors, which are the functions $f(\lambda)$ and $g(\lambda)$. We can now develop the set of $N$ SPDs in an $M$-dimensional orthonormal basis $\{\varphi_j(\lambda)\}_j$, with $\langle\, \varphi_i\, | \, \varphi_j \, \rangle = \delta_{ij}$, $\delta_{ij}$ denoting the Kronecker delta, and ${\rm 1} \le M \le N$:
\begin{equation}
s_i(\lambda) = \sum\limits_{j=1}^{M}\, a_{ij}\cdot \varphi_j(\lambda) \quad . \label{eq:2}
\end{equation}
Combining Eqs. (\ref{eq:1}) and (\ref{eq:2}) for all $N$ calibration sources, we obtain
\begin{equation}
{\bf c} = {\bf A}\, {\bf p} \quad , \label{eq:linearSystem}
\end{equation} 
with $\bf A$ the $N \times M$ matrix containing the coefficients $a_{ij}$ for source $i$ and basis function $j$, and $\bf p$ the $M$-vector containing the elements $p_j = \langle\, p \, | \, \varphi_j \, \rangle$. The values of $p_j$ are thus the projections of the passband $p(\lambda)$ onto the basis function $\varphi_j(\lambda)$, and we denote the linear combination
\begin{equation}
p_\parallel(\lambda) = \sum\limits_{j=1}^{M} \, p_j \cdot \varphi_j(\lambda)
\end{equation}
the ''parallel component'' of the passband $p(\lambda)$. This function is uniquely defined by the set of calibration sources and is obtained by solving Eq. (\ref{eq:linearSystem}) for $\bf p$. However, we have the freedom of adding any function $p_\perp(\lambda)$ that satisfies the condition
\begin{equation}
\langle \, p_\perp \, | \, \varphi_j \, \rangle = 0 \;\; \text{for all}\; i,\;\; i=1,\ldots ,M \quad
\end{equation}
to the parallel component $p_\parallel(\lambda)$ without affecting $\bf c$. The function $p_\perp(\lambda)$, which we call the ''orthogonal component'' of the passband $p(\lambda)$, is entirely unconstrained by the calibration sources and this function has to be estimated under the constraint that the sum
\begin{equation}
p(\lambda) = p_\parallel(\lambda) + p_\perp(\lambda) \label{eq:decomposition}
\end{equation}
satisfies all physical constraints that apply to the passband (i.e. non-negativity, smoothness, bound to unity) and that the sum meets the a-priori information on the passband.\par
The $M$ orthonormal basis functions $\{\varphi_j(\lambda)\}_j$ suitable for representing the SPDs of the set of $N$ calibration sources we construct using functional principal component analysis on the tabulated SPDs of the calibration sources. The number $M$ we estimate from the residuals obtained in solving Eq. (\ref{eq:linearSystem}). To estimate the orthogonal component of the passband, $p_\perp(\lambda)$, we start from an initial guess for the passband $p(\lambda)$, which we denote $p_{ini}(\lambda)$. This initial guess we modify with a linear multiplicative model, that is, we write
\begin{equation}
p(\lambda) = \left( \, \sum_{k=0}^{K-1}\, \alpha_k\, \phi_k(\lambda)\, \right) \cdot p_{ini}(\lambda) \quad . \label{eq:modificationModel}
\end{equation} 
Here, $\phi_k(\lambda)$ are some suitably but otherwise freely chosen basis functions, and the $\alpha_k$ are coefficients to be optimised. The difference to the standard approach of deriving a passband by modifying the shape of an initial guess for the passband is that we enforce the modified passband to have a parallel component that is obtained from the solution of Eq. (\ref{eq:linearSystem}), that is, we find the $\alpha_k$ under the constraint
\begin{equation}
\left\langle\, \left( \, \sum_{k=0}^{K-1}\, \alpha_k\, \phi_k(\lambda)\, \right) \cdot p_{ini} \, | \, \varphi_j\, \right\rangle = p_j \quad .
\end{equation}
With the matrix $\bf M$, $M_{n,m}=\langle \, \phi_m\, p_{ini}\, | \, \varphi_n\, \rangle$, we thus obtain the coefficients $\alpha_k$ by solving the linear equations
\begin{equation}
{\bf p} = {\bf M} \, {\boldsymbol \alpha} \quad . \label{eq:alpha}
\end{equation}
Up to here, the approach is identical to the one outlined by \cite{Weiler2018}, and the reader is referred to this work for a more detailed description. \cite{Weiler2018} take into account the uncertainty in $\bf p$ by replacing $\bf p$ with random sampled vectors generated using the formal variance-covariance matrix of the solution from Eq. (\ref{eq:linearSystem}). This procedure is referred to as the ''random sampling approach'' by \cite{Weiler2018}. Furthermore, free parameters for modifying $p(\lambda)$ without affecting $p_\parallel(\lambda)$ were introduced by making Eq. (\ref{eq:alpha}) underdetermined and computing a basis for the null space of the matrix $\bf M$. \cite{Weiler2018} refers to this as the ''null space approach''. We now introduce some small modifications that concern these two last aspects and which have proven useful for deriving {\it Gaia} DR2 passbands.\par
For the linear modification model according to Eq. (\ref{eq:modificationModel}), \cite{Weiler2018} have used a polynomial, $\phi_k(\lambda) = \lambda^k$. For the case of {\it Gaia} DR2 passbands, we found that we can use a larger number of basis functions $M$ than was used by \cite{Weiler2018}, and that a polynomial modification model does not provide sufficient flexibility for finding satisfying estimates for $p_\perp(\lambda)$. In this work we use cubic B-splines instead. By adjusting the knot sequence defining the B-spline basis functions $B_k(\lambda)$, much more flexibility in $p(\lambda)$ is obtained while preserving $p_\parallel(\lambda)$. In order to allow for the identity transformation, we include a constant function in the set of B-spline basis functions $B_k(\lambda)$, that is, $\phi_0(\lambda)\equiv 1$, $\phi_k(\lambda) = B_k(\lambda)$, $k=1,\ldots,K$. This way, we easily obtain satisfying estimates for $p(\lambda)$ for a given parallel component, in the sense that the obtained solutions for $p(\lambda)$ are sufficiently smooth, bound to the interval $[0,1]$, and close to the initial guess $p_{ini}(\lambda)$.\par
The second change with respect to \cite{Weiler2018} is that we combine the random sampling approach and the null space approach. We choose $K > M$, and compute the null space of $\bf M$; the $K \times (K-M)$ basis matrix of the null space we denote $\bf N$. We then generate random solutions ${\bf p}_{random}$ from $\bf p$, taking $\bf p$ and the variance-covariance matrix on $\bf p$ from Eq. (\ref{eq:linearSystem}). For each random solution, we choose the $K-M$-element vector of free parameters $\bf x$ such that $\boldsymbol \alpha = {\boldsymbol \alpha}_0 + {\bf N} \, {\bf x}$ minimises the difference between the modified initial guess for the passband and the initial guess passband, $\boldsymbol{\alpha}_0$ being one particular solution of the underdetermined system given by Eq.~(\ref{eq:alpha}). As a measure for this difference we take the $l_2$-norm of $\boldsymbol \alpha$, excluding the zero entry in $\boldsymbol \alpha$, corresponding to the constant function $\phi_0$, from computing the norm. Excluding the zero entry of $\boldsymbol \alpha$ from evaluating the difference makes the approach insensitive to a scaling factor between the initial guess and the modified passband. This way, after a small number of random samplings (less than 100 samples, typically), we obtain good estimates for $p(\lambda)$.\par
We complete the theoretical considerations of the problem of passband determination in this work by considering the uncertainty on the derived passband. As the passband $p(\lambda)$, according to Eq. (\ref{eq:decomposition}), is the sum of a constrained function $p_\parallel(\lambda)$ and an unconstrained function $p_\perp(\lambda)$, the error on the passband $p(\lambda)$ is not a well defined concept. An error on $p_\parallel(\lambda)$ can in principle be obtained, in the form of a variance-covariance matrix on the coefficient vector $\bf p$. However, as the orthogonal component $p_\perp(\lambda)$ can be chosen arbitrarily, there is no error on this component, and consequently no meaningful error on the sum of $p_\perp(\lambda)$ and $p_\parallel(\lambda)$. Modifying an initial passband with some version of a modification model as in \cite{Evans2018} or \cite{Weiler2018} may allow us to compute an error on the coefficients of that particular model. However, choosing another initial passband or modification model will result in a solution for $p(\lambda)$ that predicts the same synthetic photometry for all calibration sources, but has a different confidence interval. This effect can be seen in the two sets of passbands, the {\it DR2} and the {\it REV} passbands, by \cite{Evans2018}. Both sets of passbands result essentially in the same residuals on the photometry of the calibration sources, indicating that they both have the same parallel component with respect to the SPDs of the calibration sources used. The strong differences in shape between the {\it DR2} and {\it REV} passbands only affect orthogonal components, which are not constrained by the calibration sources. Although the residuals obtained with the two different sets of passbands are virtually the same, the error intervals specified by \cite{Evans2018} strongly deviate from each other in some wavelength ranges, indicating the model-dependency of the error intervals. The uncertainty on $p(\lambda)$ is therefore as arbitrary as the choice of the modification model, using a multiplication with polynomials or B-spline basis functions or adding principal components resulting from a simulated data set or whatever else. When comparing the passbands derived in this work with the passbands by \cite{Evans2018}, we therefore do not consider the error intervals provided. As the error on $p_\parallel(\lambda)$ is dominated by the error in the calibration spectra, for which no suitable error model is currently available \citep{Weiler2018}, we also do not provide errors on the parallel component in this work.

\subsection{Colour-colour relations \label{sec:colourcolourtheory}}

For sources with an SPD that lies within the subspace of ${\mathcal L}^2(I)$ spanned by the SPDs of the calibrations sources (that is, for SPDs that can be well approximated by a linear combination of the SPDs of the calibration sources), the synthetic photometry depends on the constrained parallel component of the passband. For sources with a significant component in their SPD that falls outside the space spanned by the calibration sources, the synthetic photometry also depends on the guess for the orthogonal passband component. The larger the component of an SPD outside the subspace spanned by the calibration sources is in comparison to its component inside the subspace, the stronger the contribution of the unconstrained passband component to the synthetic photometry becomes. This effect affects objects with very non-stellar SPDs, such as quasars, and, as already demonstrated in \cite{Weiler2018}, very red sources not included in the set of calibration sources. An incorrect estimate of $p_\perp(\lambda)$ can thus introduce systematic errors in the synthetic photometry for such sources. In the case of very red sources, the systematic error increases with decreasing effective temperature of the objects.\par
To reduce the effects of such systematic errors, we make use of empirical spectra representative for different spectral types, as well as theoretical stellar spectra, including spectral types not included in the set of calibration sources. For these spectra, no individual astrophysical source is available for comparing synthetic and {\it Gaia} DR2 photometry. Instead we compare the synthetic photometry in a statistical way with observations, by comparing the path the empirical and theoretical spectra follow in a colour-colour diagram with the curve the observed photometry of a large set of sources forms in the same diagram. We confirm that, for passbands derived from spectral libraries as described in Sect.~\ref{sec:math}, the path in the colour-colour diagram for empirical and theoretical spectra agrees well with the observational relation for sources of spectral types similar to the calibration sources. We then assume that the agreement between the empirical and theoretical and observational colour-colour relations should also hold for sources with significant components of their SPDs outside the space spanned by the SPDs of the calibration sources. We thus add the additional constraint on the orthogonal component that it should not only result from a smooth variation of the initial guess for the passband, but at the same time also result in a good agreement between the predicted and observed colour-colour relation for stars of all spectral types, including very red sources.\par
To find such an estimate for $p_\perp(\lambda)$, we employ the random sampling introduced in Sec.~\ref{sec:math}, generating candidates for $p_\perp(\lambda)$ that result in smooth and physically reasonable passbands $p(\lambda) = p_\parallel(\lambda) + p_\perp(\lambda)$ randomly, and comparing the synthetic colour-colour relation for the empirical and theoretical spectra with the one observed for {\it Gaia} DR2. In practice, we first determine the $G$ passband. We find this passband in good agreement with the pre-launch expectations, and therefore do not subject it to further modifications by changing its orthogonal component. In the next step we determine the $BP$ passband. We compare the synthetic {\it BP--RP} versus {\it BP--G} colours with the observed distribution, and find a good agreement for all empirical and  theoretical spectra similar to the calibration spectra. Finally, we determine the {\it RP} passband and refine it by selecting an orthogonal component that results in a good agreement between the empirical and theoretical spectra and the observational relation in the $BP-G$ versus $G-RP$ diagram for all spectral types.

   \begin{figure*}
   \centering
   \includegraphics[width=0.98\textwidth]{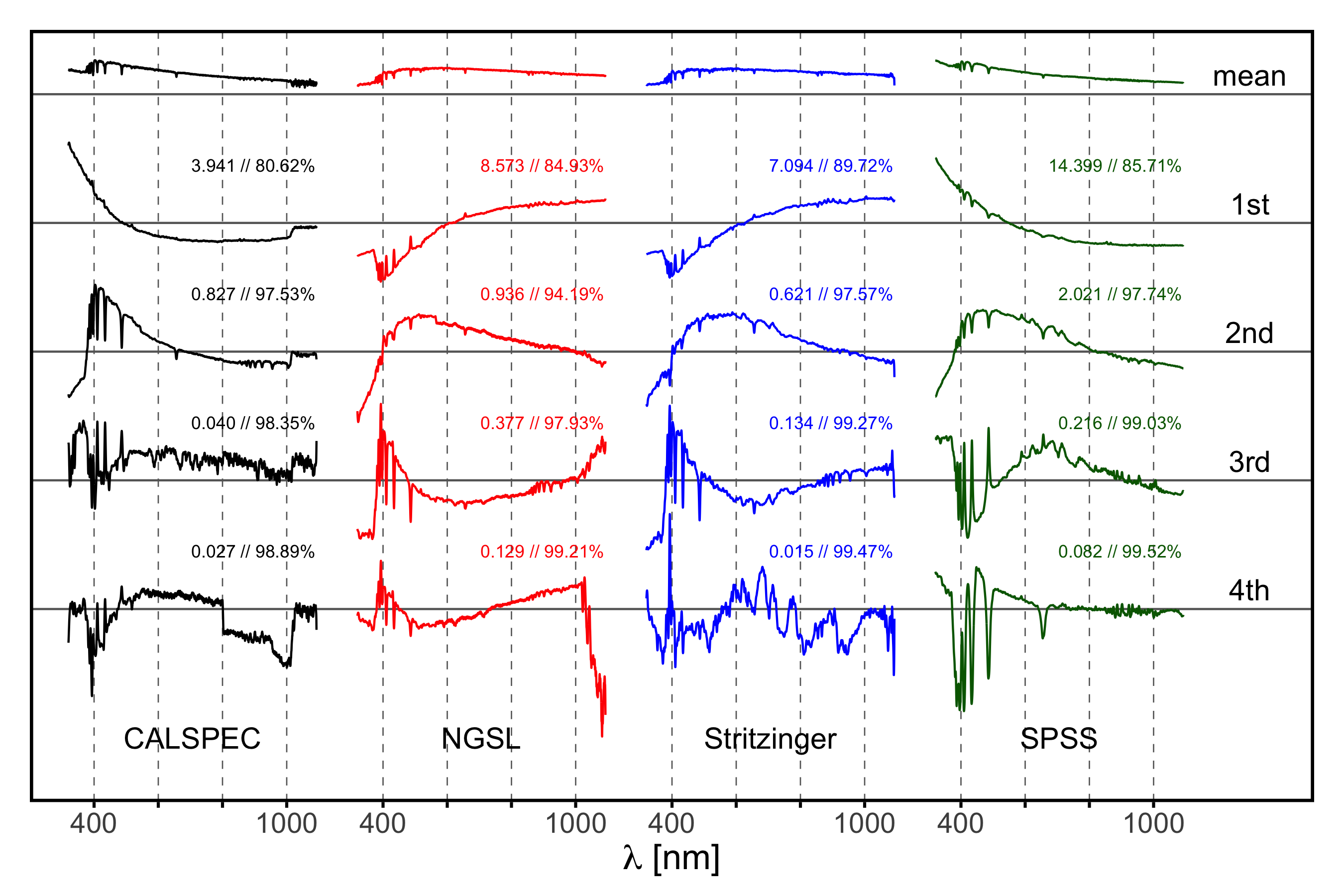}
   \caption{Mean functions and first four functional principal components for the four different spectral libraries. All functions are plotted in the same scale and the eigenfunctions are normalised with respect to the $l_2$-norm. The numbers give the associated eigenvalues and the cumulative fraction of variance explained (FVE).}
              \label{Fig:decomposition}
    \end{figure*}

\subsection{Passband parameters \label{sec:passbandparameters}}

For each passband derived in this work, we provide a set of relevant parameters. These parameters include the following.
\begin{itemize}
\item The zero point in the VEGAMAG photometric system. This photometric system uses the SED of Vega to compute the reference flux in each particular passband. For the SED of Vega, we use the spectrum {\tt alpha\_lyr\_stis\_008}, the latest CALSPEC spectrum for Vega, and we assume a magnitude of 0.023 for Vega \citep{Bohlin2007}.\\
\item The zero point in the AB photometric system \citep{OkeGunn}. This photometric system uses a constant SED of 3631~Jy at all frequencies to compute the reference flux in particular passbands.\\
\item The mean wavelength $\lambda_m$ of the passband, defined as the weighted mean
\begin{equation}
\lambda_m \equiv \frac{\int_{\lambda_0}^{\lambda_1}\, \lambda \cdot p(\lambda)\, {\rm d}\lambda} {\int_{\lambda_0}^{\lambda_1}\, p(\lambda)\,{\rm d}\lambda} \quad .
\end{equation}
\item The pivot wavelength $\lambda_p$ \citep{Koornneef}, defined as
\begin{equation}
\lambda_p \equiv \sqrt{\frac{\int_{\lambda_0}^{\lambda_1}\, \lambda \cdot p(\lambda)\,{\rm d}\lambda}{\int_{\lambda_0}^{\lambda_1}\, \lambda^{-1}\cdot p(\lambda)\,{\rm d}\lambda}} \quad .
\end{equation}
This quantity is convenient for conversion between fluxes expressed in frequencies and in wavelength.\\
\item The $l_2$-norms of the parallel and orthogonal components of the passbands, $||\, p_\parallel \,||_2$ and $||\, p_\perp \, ||_2$. These quantities are required to compute the angle $\gamma$ defined as \citep{Weiler2018}
 \begin{equation}
\gamma = {\rm atan}\left( \,\left| \,  \frac{\langle\, p_\perp\,|\, s\, \rangle}{||p_\perp ||_2} \frac{||\,p_\parallel\,||_2}{\langle\, p_\parallel\,|\,s\, \rangle} \, \right| \, \right) \quad , \label{eq:gamma}
\end{equation}
which serves as a measure of how well a given SPD is represented by the SPDs of the calibration spectra used in the derivation of the passbands. A value of $\gamma$ of zero corresponds to a source whose synthetic photometry does not depend on the choice of $p_\perp(\lambda)$, while a value of 90$^\circ$ corresponds to a source whose synthetic photometry does not depend on $p_\parallel(\lambda)$. It therefore serves as an indicator of how much the synthetic photometry of a given SPD depends on the choice of $p_\perp(\lambda)$, and with it how sensitive it is to systematic errors resulting from an incorrect estimate for $p_\perp(\lambda)$.
\end{itemize}
The values of these parameters are summarised in Table~\ref{tab:1} for all passbands derived in this work. All passbands are tabulated with their parallel and orthogonal components in Table~2, which is available in electronic form only.

\section{Calibration data \label{sec:data}}

\subsection{Observational spectral libraries \label{sec:obslibraries}}

In this work, we consider four spectral libraries for calibration sources, namely the CALSPEC library \citep{Bohlin2017}, the {\it Next Generation Spectral Library} (NGSL) \citep{HeapLindler}, the spectral library by \cite{Stritzinger2005}, and the set of spectrophotometric standard stars by \cite{Elena2012}.\par
The CALSPEC set consists of calibration sources for the instruments of the Hubble Space Telescope (HST), and provides a particularly wide coverage in magnitudes, from very bright down to about 17. The set of spectra selected for this work contains sources of spectral type O to K that fall in the suitable magnitude range to be used for {\it Gaia} DR2 calibration. The NGSL library is based on HST/Space Telescope Imaging Spectrograph (STIS) observations and covers $G$ magnitudes from very bright down to about 12, and also contains sources of the spectral types O to K. The wavelength coverage is from the ultraviolet to about 1030 nm, and we applied the same extension to 1100 nm as discussed in \cite{Weiler2018}. The \cite{Stritzinger2005} set of spectra are obtained from ground-based observations, covering intermediate magnitudes roughly between 7 and 12 in $G$. This data set contains spectra of spectral types B to K, plus one M-type source.\par
To calibrate {\it Gaia} data, a homogeneous set of approximately 200 spectrophotometric standard stars (SPSS) is being built \citep{Elena2012,Altavilla2015}, covering the widest possible range in spectral types and the entire {\it Gaia} wavelength range (330~nm to 1050~nm), and calibrated on Vega \citep{Bohlin2004,Bohlin2007,Bohlin2014} with errors within 1\%\ to 3\%. The SPSS set was also monitored for constancy within 10~mmag \citep{Marinoni2016}. Because the grid of calibration spectra is not yet complete and published, for this work we use the first post-{\it Gaia}-launch internally released version\footnote{Kindly provided by E.~Pancino and the SPSS DPAC team.}, which was also used to calibrate the {\it Gaia}~DR2 photometric data by \cite{Evans2018}. More information on the SPSS can be found in the {\it Gaia}~DR2 online documentation, Chapter 5: Photometry. The SPSS  cover the $G$ magnitude range from about nine to 15. It contains calibration sources of spectral types O to K, including many white dwarfs. Additionally, the SPSS include several calibration sources of spectral type M1V and M2V, thus providing good means for an improved calibration for very red sources.\par
We only consider sources from these libraries that are not marked as variable in either the Simbad data base or the HIPPARCOS catalogue. For NGSL we exclude also stars for which no slit throughput correction could be applied. For the comparison with {\it Gaia} DR2 photometry we introduce a lower limit in magnitude. As reported by \cite{Evans2018}, bright sources are affected by saturation effects. To exclude such effects from affecting the determination of the passbands, we restrict our calibration sources to $G$ magnitudes larger than 5.9. For brighter objects strong systematic effects in $G$ residuals due to saturation are detectible. For $BP$ and $RP$, the dispersion of {\it Gaia} observations allows for the observation of brighter objects without notable saturation effects, and we chose a lower magnitude limit of 5.0. A slightly lower limit may still be possible, but to rule out saturation effects even for sources with extreme colours, we chose the conservative limit of 5.0.\par
The selected stars were crossmatched with the {\it Gaia}~DR2, resulting in a set of 45 CALSPEC sources, 210 NGSL sources, 72 Stritzinger sources, and 92 SPSS. Functional principal component analysis was then applied to each of these four sets of spectra independently to construct four sets of basis functions $\{ \varphi_j(\lambda) \}_j$, analogous to \cite{Weiler2018}. The mean function and the first four functional principal components are shown in Fig.~\ref{Fig:decomposition} for the four data sets for illustration. The principal components are normalised with respect to the $l_2$-norm, and the corresponding eigenvalues, displayed in Fig.~\ref{Fig:decomposition}, describe the weight each component has in describing the entire data set. Figure~\ref{Fig:decomposition} also includes the cumulative fraction of variance explained (FVE), indicating the rather low dimensionality of the data set. With a linear combination of four principal components, a 99\%-level in the capture of the variance of the spectral data sets is achieved.\par
The principal components show typical features of stellar spectra. However, they do not have any physical meaning in themselves, but rather provide a compact empirical description of the set of stellar spectra used to generate them. They thus depend on the underlying set of stellar spectra, which is different for the different spectral libraries, resulting in different functions. There are also features visible in the principal components that are most likely not of astrophysical origin, such as abrupt ''jumps'' with wavelength in some of the principal components. Prominent features of this kind can be seen in the fourth principal component of the CALSPEC set, which shows a clear displacement between about 801.5~nm and 1015.5~nm, and a less prominent one at 565~nm in the second principal component of the NGSL set. These discontinuities result from combining spectra on different wavelength intervals into a single one. In the CALSPEC set, the discontinuity at 801.5~nm results from a single spectrum, for which the combination of spectral data on two wavelength intervals results in a strong abrupt change in the noise level. This change results in an artefact in the fourth principal component. The broader feature at around 1015~nm in the CALSPEC data set is also an artefact from the combination of different data to a single spectrum, but the combination occurs for many spectra and at slightly different wavelengths. As a consequence this artefact in the functional principal components is less localised in wavelength. The discontinuity in the second principal component for the NGSL data set is caused by a small discontinuity when combining spectra obtained with two different gratings to a single one. This effect is in fact very small in any individual spectrum, but it occurs systematically and is therefore already included in the second principal component. Although the discontinuities in the principal components have little impact on the SPD of any particular source, as the projection of the SPD onto these principal components is small, the projection of the passband onto these principal components may not be small. The artefacts may therefore become amplified and result in unphysical parallel components $p_\parallel(\lambda)$. These features do not pose difficulties in the estimation of the passband $p(\lambda)$, as their effect can easily be compensated for by a suitable choice of the orthogonal component $p_\perp(\lambda)$. The separation into parallel and orthogonal component, however, may become flawed. We therefore seek to avoid discontinuities in the passband determination. The SPSS data set is free from such effects, and together with the wide coverage in different spectral types, we consider the SPSS set of calibration sources the most suitable for the determination of the shape of the {\it Gaia} passbands. We therefore do not combine the different spectral libraries, but rather use the SPSS library as the ''default'' set of calibration sources in the passband determination, and validate the results obtained with this data set by ensuring that the synthetic magnitudes for the remaining three sets of calibration sources are in good agreement with the observed magnitudes as well.

\subsection{Empirical and theoretical spectral libraries}

As a set of empirical spectra representative of different spectral types, we use the \cite{Pickles} library of stellar spectra. This library includes SEDs for all spectral types from O5 to M10, and for luminosity classes I to V, which were derived from observational spectra. Different metallicities are also included for certain spectral types. \cite{Pickles} presents two sets of SEDs, covering the spectral ranges from 115~nm to 1062~nm (the UVILIB library) and from 115~nm to 2.5~$\mu$m (the UVKLIB library). In this work, we use both.\par
As a second set of spectra, we make use of the BaSeL 3.1 WLBC99 library of stellar spectra \citep{Westera2002}. This library provides theoretical spectra on a wide grid of effective temperatures, surface gravities, and metallicities. We refer to the colour-colour relations of both the empirical library by \cite{Pickles} and the theoretical BaSeL library as the synthetic colour-colour relations. For the optimisation of the passbands, we only use the \cite{Pickles} spectra in order to avoid an optimisation of the passbands with respect to any particular astrophysical parameters in the BaSeL library. The BaSeL spectra are rather considered as a test case for the result obtained with the Pickles spectra.

\subsection{\textit{Gaia} DR2 data \label{sec:dr2}}

In order to compare synthetic colour-colour relationships with observational ones, we select a set of {\it Gaia} DR2 photometric observations. Several issues in the {\it Gaia} DR2 photometry that could affect the passband determination have been reported by \cite{Evans2018} and \cite{Arenou2018}. When selecting {\it Gaia} DR2 observations for our passband determinations, we avoid these issues by applying filters in sky region and in magnitude. To reduce effects caused by crowding, we exclude the region around the galactic plane within $\pm$ 30$^\circ$ galactic latitude. To minimise effects introduced by inaccurate background correction in the {\it Gaia} DR2 calibration, we restrict the data set to relatively bright sources, with $BP$ and $RP$ magnitudes less than 17. To exclude the effects of saturation, as mentioned in Sect.~\ref{sec:obslibraries}, we furthermore introduce a lower magnitude limit of six in $G$ and five in $BP$ and $RP$. Furthermore, we only include sources for which at least ten photometric observations in each band entered in the computation of the {\it Gaia} DR2 mean photometry.\par
\begin{figure*}
   \centering
   \begin{minipage}[t]{0.99\textwidth}
   \includegraphics[width=\textwidth]{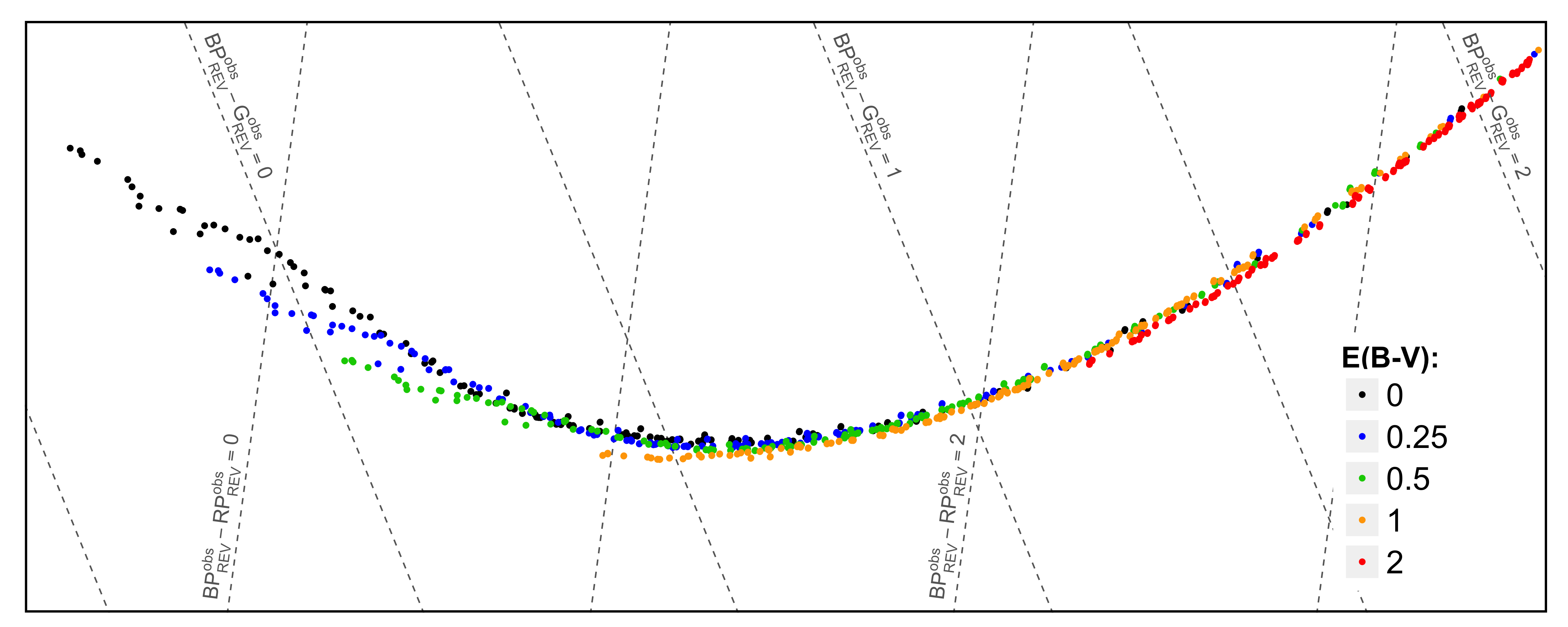}
   \end{minipage}
   \begin{minipage}[l]{0.5\textwidth}
   \includegraphics[width=\textwidth]{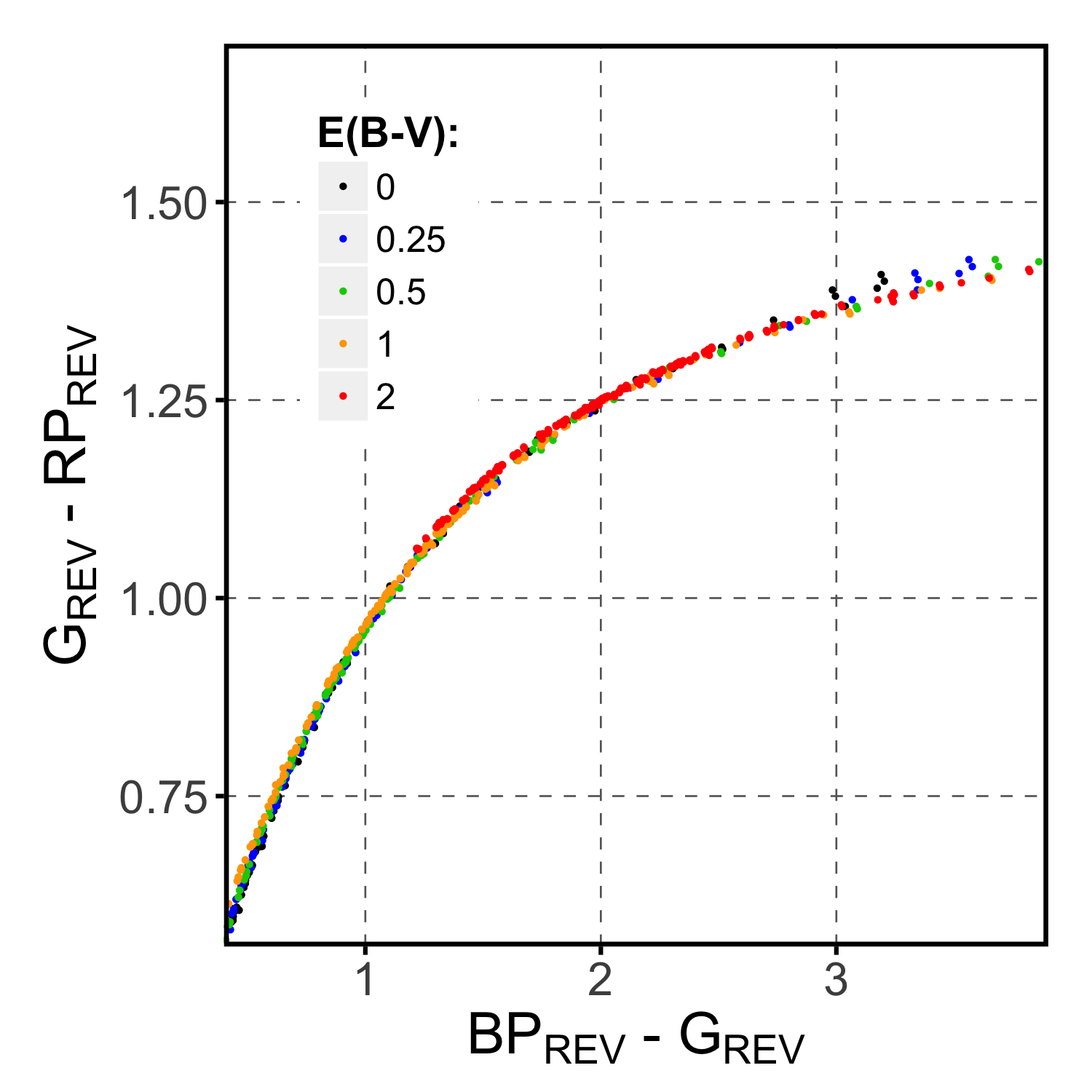}
   \end{minipage}
   \begin{minipage}[t]{0.4\textwidth}
   \caption{Position of stellar spectra by \cite{Pickles} in colour-colour diagrams for different levels of interstellar reddening. Top panel: The region of the $BP-RP$ versus $BP-G$ diagram considered in Sect.~\ref{sec:BP}. The colour-colour diagram is rotated clockwise by 30$^\circ$ and exaggerated in the vertical direction by a factor 11. The dashed lines correspond to constant $BP-RP$ and $BP-G$ values. Left panel: The red-end region of the $BP-G$ versus $G-RP$ diagram used in Sect.~\ref{sec:RP}.}
              \label{Fig:extinction}
        \end{minipage}
    \end{figure*}

Excluding the galactic plane from our {\it Gaia} DR2 data set also avoids sources whose SPDs are extremely reddened by interstellar extinction. In order to estimate the influence of interstellar extinction on the colour-colour relations used in this work, we applied the extinction law by \cite{Cardelli1989} to the Pickles spectra, and computed the colour-colour relations resulting for different colour excesses $E(B-V)$ in the regions of the colour-colour diagrams studied in this work. The results are shown in Fig.~\ref{Fig:extinction}. For the $BP - RP$ versus $BP - G$ case, stars follow a relatively narrow path running close to diagonal through the diagram. To avoid an inconveniently large figure with many white areas, we therefore rotate the diagram by 30$^\circ$ for a more convenient fit of the path within a rectangular plotting region. The main effect of extinction is a shift along the path of the un-reddened spectra in the colour-colour diagram, with only small displacements away from this path. The conclusions in this work drawn from comparisons between synthetic and observational colour-colour relations will therefore not be influenced by interstellar reddening.

\section{\textit{G} passband \label{sec:G}}

\subsection{Determination of the passband}

\begin{figure*}
   \centering
   \includegraphics[width=0.48\textwidth]{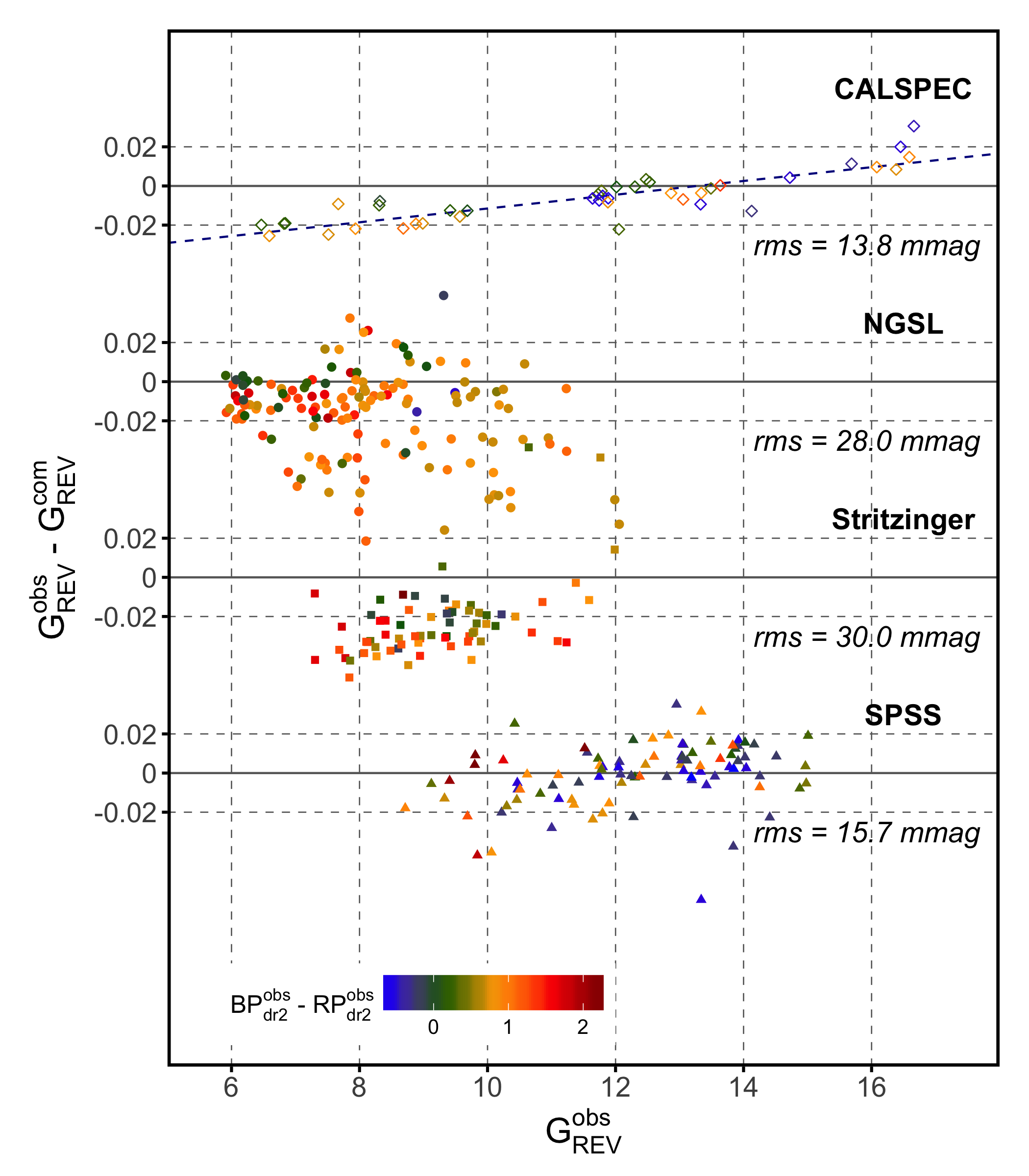}
   \includegraphics[width=0.48\textwidth]{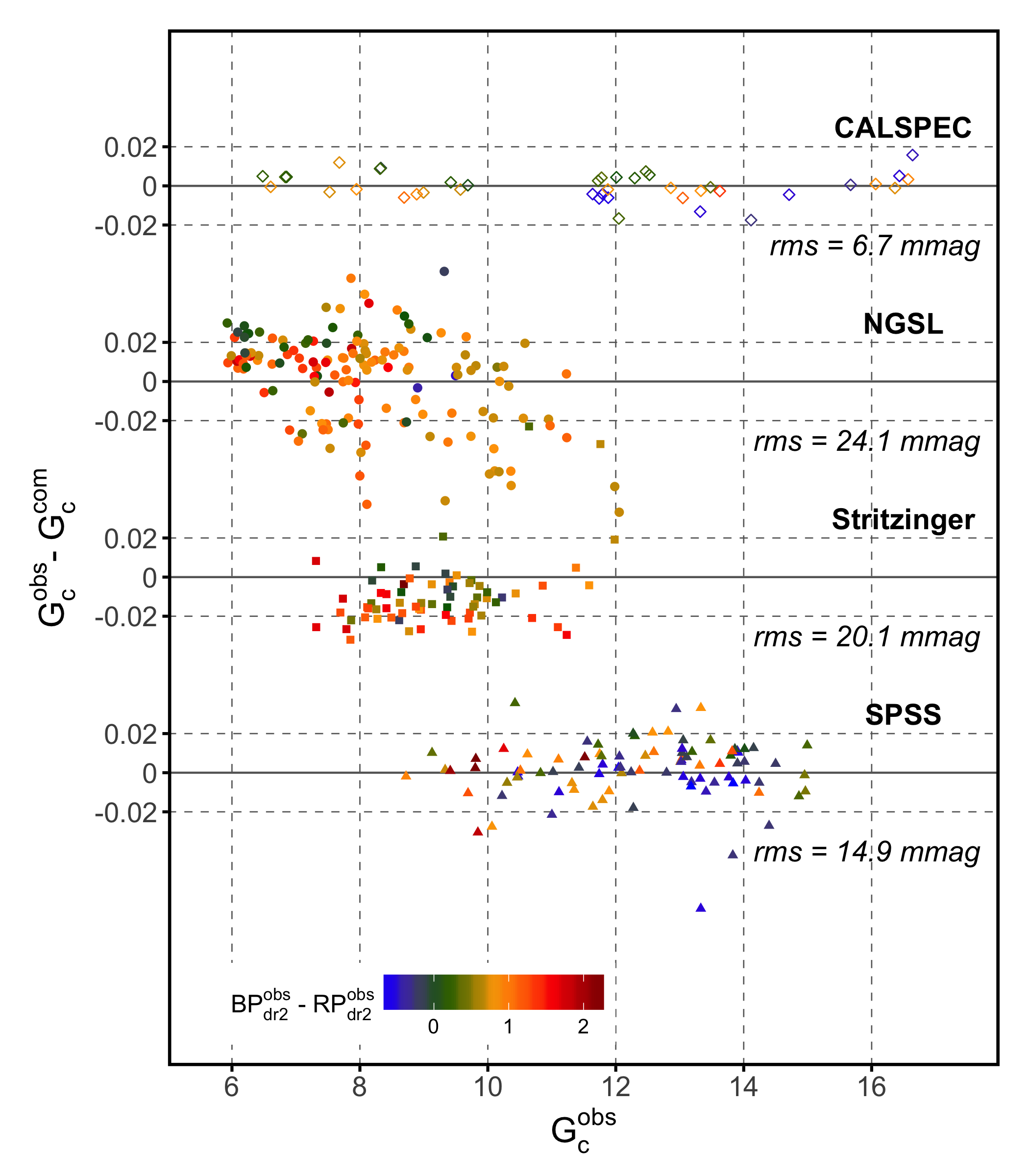}
   \caption{Residuals for the four libraries used in this work as a function of magnitude for $G$ for the {\it REV} passband (left) and the passband derived in this work, including a linear correction term (right). The colour coding indicates the value of $BP-RP$. The rms of the residuals for the four data sets are listed.}
              \label{Fig:residualsG}
    \end{figure*}

Using the $G$ passbands from \cite{Evans2018}, one obtains the residuals for the four spectral libraries shown in Fig.~\ref{Fig:residualsG}, left panel, plotted versus magnitude. The $BP-RP$ colour of the sources is indicated by the colour coding. This plot differs from the corresponding plot in \cite{Evans2018} only because of the different criteria in the selection of sources that enter into the plot, and it is repeated here for easy direct comparison. The root mean square (rms) values of the residuals for the four different sets of spectra are included in the figure. The dominant feature in the residuals is a magnitude dependency, which is most clearly visible for the  CALSPEC data set, as it has a low random noise and covers the widest range of magnitudes. A systematic trend in CALSPEC residuals with magnitude has already been mentioned by \cite{Evans2018}, and \cite{Arenou2018} reported a trend in the $G - BP$ residuals with $G$ magnitude in the order of a few milli-magnitudes per magnitude. We thus confirm this trend within a range of $G$ magnitudes between about six and 17. Using the CALSPEC data set, we find it well approximated with a colour-independent linear correction of $\rm 3.5$~mmag/magnitude, with an uncertainty of about $\pm$0.3~mmag/magnitude. As the correction is on the level of a few tens of a milli-magnitude, a linear correction in magnitude and a linear correction in flux is virtually equivalent, and we choose a correction in magnitude for its more convenient formulation in magnitude systems. An approximately linear drift in the $G$ photometric system with magnitude was already reported for the {\it Gaia} Data Release~1 \citep{Weiler2018} and the reason for this remains unclear for the time being.\par
The drift in the $G$ magnitude may affect the determination of the passband, even if independent from the colour, as the calibration sources have a non-uniform distribution of colours with apparent magnitude, the red sources being in tendency brighter than the blue sources. We therefore correct the $G$ photometry by introducing a linear correction term in magnitude before determining the $G$ passband, that is, we assume a relationship of
\begin{equation}
G_c = 0.9965 \cdot (-2.5)\cdot {\rm log_{10}}\left( I_G \right) + zp \label{eq:Gmag}
\end{equation}
between $I_G$, the observed counts in the $G$ passband (in electrons per second within the {\it Gaia} aperture), and the $G_c$ magnitude. The factor of 0.9965 compensates for a linear tendency in magnitude corresponding to a fading of $\rm 3.5$ mmag per magnitude and $zp$ denotes the zero-point of the passband, derived as described in Sect.~\ref{sec:passbandparameters}. For synthetic photometry in the $G$ band, the factor 0.9965 is not required, as only the observational magnitudes have to be corrected for the observed trend.\par
For the determination of $p_\parallel(\lambda)$ we use the SPSS data set with $M=5$ basis functions. We solve Eq.~(\ref{eq:linearSystem}) to obtain $p_\parallel(\lambda)$, and then estimate the full passband $p(\lambda)$ as outlined in Sect.~\ref{sec:math}. As the initial guess, $p_{ini}(\lambda),$ we use the $G$ passband by \cite{Jordi2010}. The residuals obtained with the passband solution derived in this work are shown in Fig.~\ref{Fig:residualsG}, right panel, together with the rms values of the residuals. The passband solution of this work is shown in Fig.~\ref{Fig:passbandG}, with the sum of the parallel and orthogonal component in the upper panel and the parallel component in the lower panel. Table \ref{tab:1} includes the passband parameters such as the zero points in the VEGAMAG and AB system, the mean and pivot wavelength, and the $l_2$-norms of the parallel and orthogonal component. The $G$ passband derived in this work is tabulated in Table~2, separated into its parallel and orthogonal components.

\subsection{Comparison with other results}

\begin{figure}
   \centering
   \includegraphics[width=0.48\textwidth]{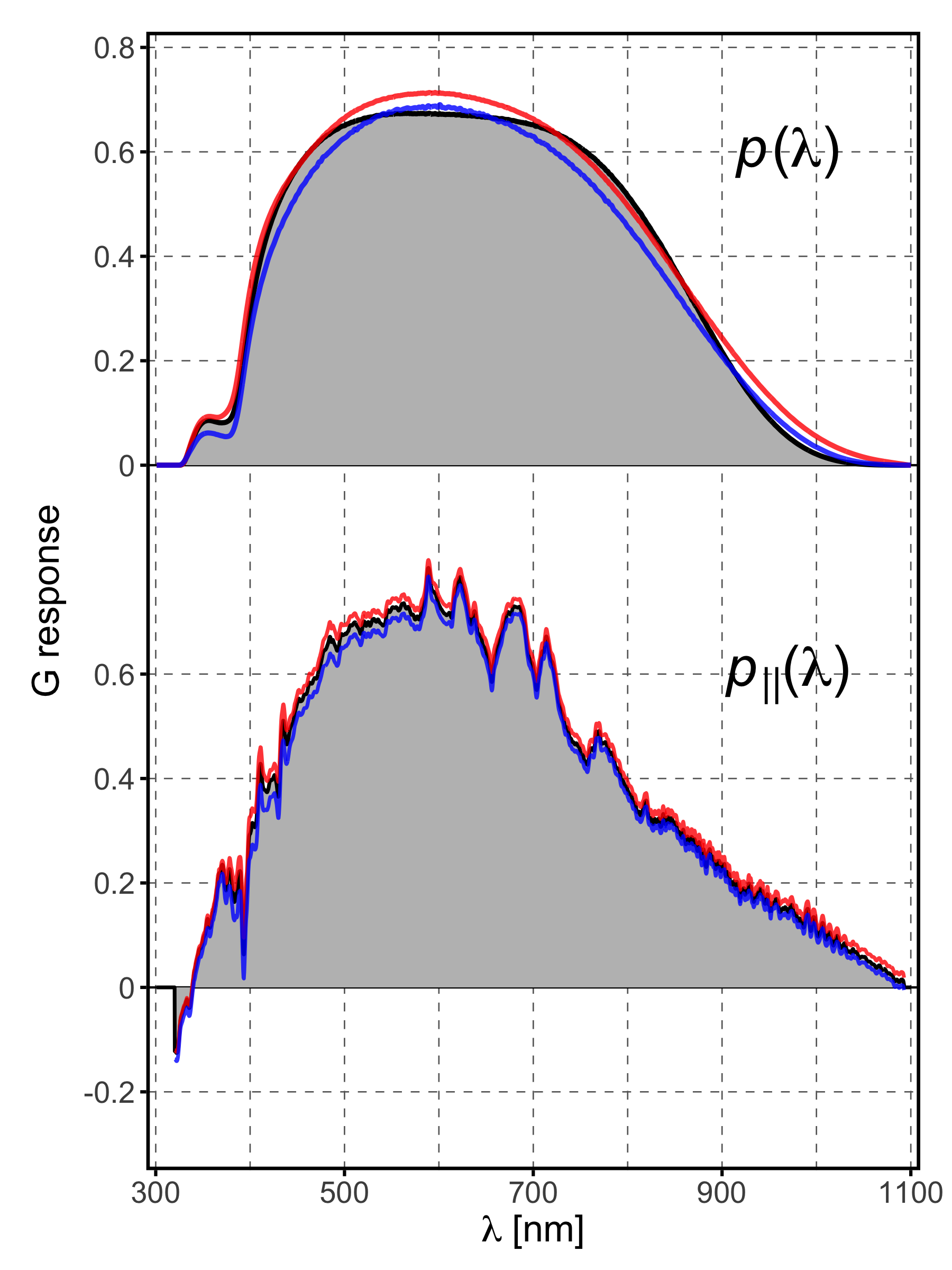}
   \caption{The solution for the $G$ passband (upper panel) and the parallel component (lower panel). Blue line: \cite{Jordi2010}. Red line: \cite{Evans2018} ({\it REV}), black line shaded: This work.}
              \label{Fig:passbandG}
    \end{figure}

The passband obtained in this work is similar to the one derived by \cite{Evans2018} and presented by \cite{Jordi2010}, which is also presented in Fig.~\ref{Fig:passbandG} for comparison. The solution of this work is slightly steeper at long wavelengths and flatter at intermediate wavelengths. Comparing the parallel component, the result of this work is again similar to the $G$ passbands by \cite{Evans2018} and \cite{Jordi2010}. Given the uncertainties estimated from the residuals, all three $G$ passband solutions are essentially equivalent. We thus find the $G$ passband in close agreement with the pre-launch expectation and the solution presented by \cite{Evans2018}. This is in contrast to the results for the {\it Gaia} Data Release 1, where significant deviations from the pre-launch expectation were reported by \cite{Maiz2017} and \cite{Weiler2018}. The deviations from the pre-launch expectation was explained by the effect of contamination of optical surfaces in the {\it Gaia} instruments present in the early stages of data collection. The current result may therefore be an indicator of reduced contamination and of an improvement in the instrumental calibration. The major effect for the $G$ passband achieved in this work results from the linear correction of the tendency observed in $G$ magnitudes. The correction results in a reduction of the rms of the residuals for all four sets of spectra considered. As the improvement is achieved by correcting a systematic effect, the change in rms, however, depends on the sources considered, and the quantitative improvement cannot be generalised.

\section{\textit{BP} passband \label{sec:BP}}

\subsection{Determination of the passband}

\begin{figure*}
   \centering
   \includegraphics[width=0.48\textwidth]{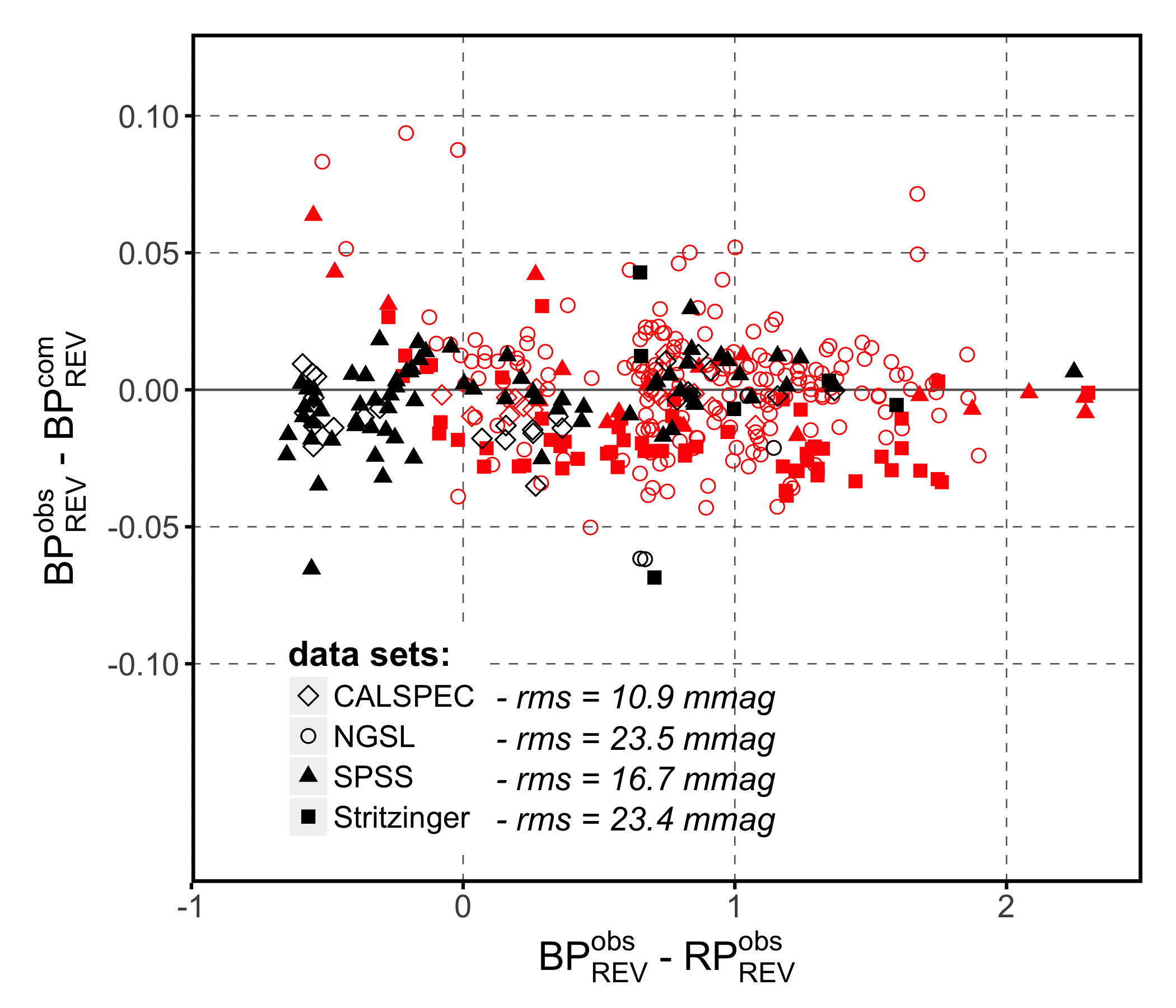}
   \includegraphics[width=0.48\textwidth]{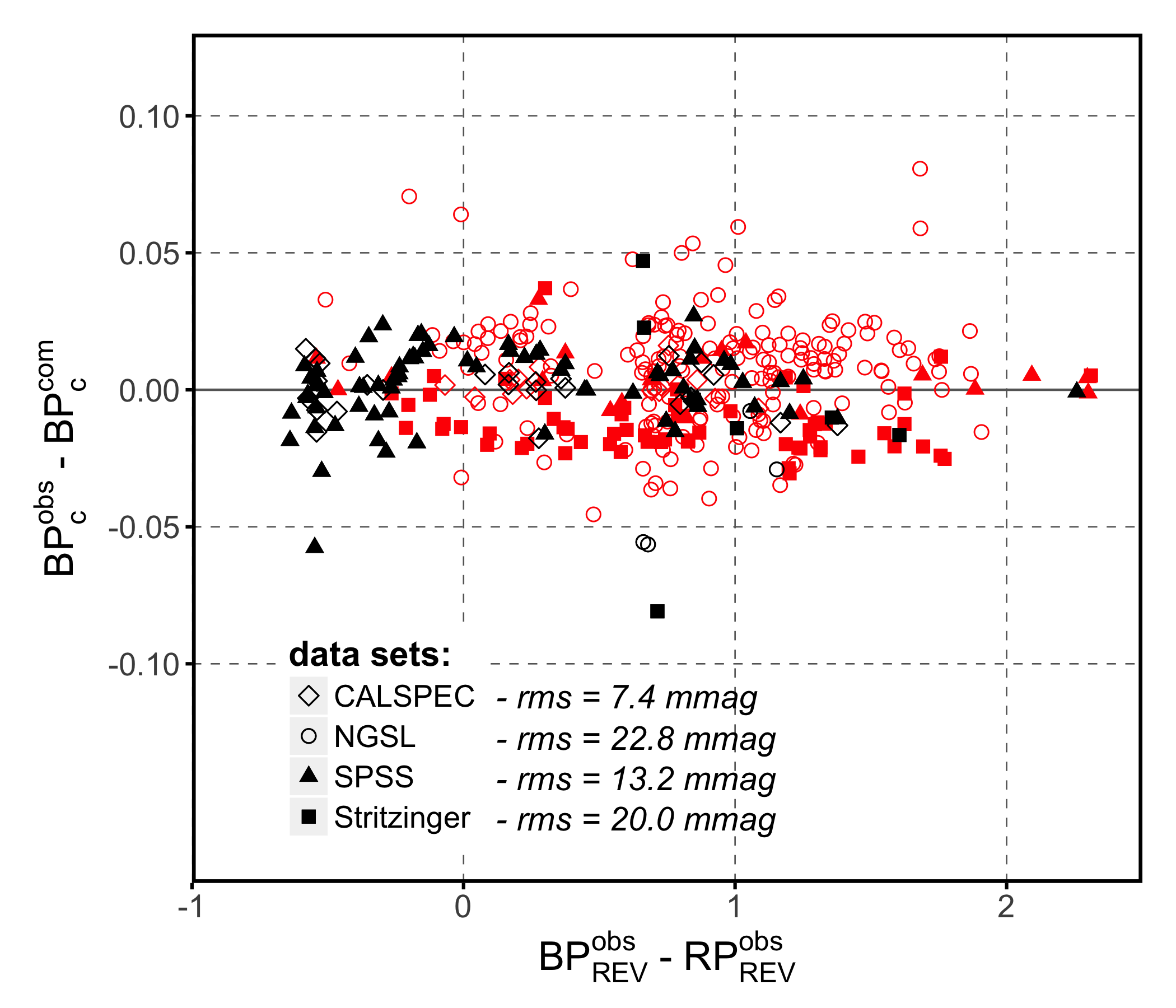}
   \caption{Residuals for the BP passband for four spectral libraries. Left panel: \cite{Evans2018}, {\it REV} passband. Right panel: This work. Residuals for sources with $G_{dr2} > {\rm 10.99}$ are plotted in black, sources with $G_{dr2} < {\rm 10.99}$ in red. The rms of the residuals are listed for the four data sets.}
              \label{Fig:residualsBP}
    \end{figure*}

The residuals for the $BP$ passband resulting from the use of the {\it REV} passband by \cite{Evans2018} are shown in the left panel of Fig.~\ref{Fig:residualsBP}. The residuals are plotted versus {\it BP -- RP} colour and grouped into two sets by $G_{dr2}$ magnitude. Sources fainter than 10.99 mag and brighter than 10.99~mag are plotted in different colours. A colour-dependent systematic behaviour in the residuals can be spotted, with blue sources brighter than 10.99 mag in $G$ band systematically rising upwards, while the residuals for the fainter sources slightly decreasing with decreasing {\it BP -- RP} colour index. This separation, visible from a {\it BP -- RP} colour of about zero downwards, is consistent with all four sets of calibration sources used in this work. For the CALSPEC data set, all sources in the colour range of {\it BP -- RP} less than zero belong to the faint group, and their residuals all belong to the lower group. For the NGSL and Stritzinger sets, all sources in the blue part belong to the bright group, and all their residuals show the increasing tendency with decreasing colour index. For the SPSS, most sources in the blue range belong to the faint $G$ magnitude range, and the corresponding residuals fall within the lower group. Three of the SPSS with {\it BP -- RP,} however, belong to the bright group with $G_{dr2}$ < 10.99, and their residuals also follow the increasing trend. The precise location of the magnitude break cannot be determined within this work. For the calibration spectra available, the optimal separation occurs within the range of 10.47 and 10.99 in $G_{dr2}$. Depending on whether we admit an ''outlier'' from the bright or faint magnitude regime, even larger or smaller values are possible. Here, we adopt the value of $G_{dr2}$ = 10.99 as an estimate for the point of separation between the two regimes.\par
The observed behaviour can be explained by the sources fainter than 10.99 in $G_{dr2}$ not being in the same photometric system as the brighter sources. This inconsistency mostly affects very blue sources with a {\it BP -- RP} colour index less than zero. Such an inconsistency may result from sources observed under specific instrumental configurations failing to converge to a common photometric system during the calibration process, since bright sources are rarer than faint sources while the complexity of the instrument that needs to be taken into consideration in the calibration process is the same for sources of all magnitudes. Extremely blue sources are also rare. As a consequence, it might be difficult to precisely link the photometric system of bright blue sources with the photometric system of faint sources if they are observed in different instrumental configurations. A possible change in instrumental configuration for sources in the observed magnitude range may be a change in CCD gate activation, a mechanism that is used by the {\it Gaia} instruments to adjust the effective exposure time for brighter sources \citep{Carrasco2016}. Another change in instrument configuration is the transition between two {\it Gaia} window classes, resulting in one-dimensional and two-dimensional spectra, respectively. These spectra are the basis for deriving the $BP$ and $RP$ magnitudes, and the change in window class is expected around $G \approx {\rm 11.5}$ \citep{Carrasco2016}. The exact reason for the discrepancy between blue sources in the two different magnitude regimes, however, remains unknown. Changes in the instrumental configuration, however, are driven by the on-board estimate of the $G$ magnitude, which is the reason why we specify the location of inconsistency in the {\it BP} photometric system by the {\it G} magnitude.\par
To account for this feature in {\it BP} magnitude, we separate the range of {\it G} magnitudes into two regimes, the ''bright'' range for sources with $G_{dr2}$ < 10.99 and the ''faint'' range for sources with $G_{dr2}$ > 10.99. We then derive the {\it BP} passband separately in the two magnitude ranges. We want to use only one set of calibration sources in both magnitude regimes in order to ensure a consistent solution between both. Only the SPSS and the CALSPEC libraries cover both regimes, and for the bright regime the number of sources is rather small. The CALSPEC data set provides a lower random noise for the bright spectra than the SPSS, and we prefer to use the CALSPEC spectra in this particular case. This set provides similar numbers of sources in the ''bright'' and ''faint'' regime, although with different coverage in spectral types. To compensate for this effect, we adjust the estimate for the orthogonal components of the two {\it BP} passbands such that a good result is achieved for the sources in the other three sets of spectra as well.\par
\begin{figure*}
   \centering
   \includegraphics[width=0.48\textwidth]{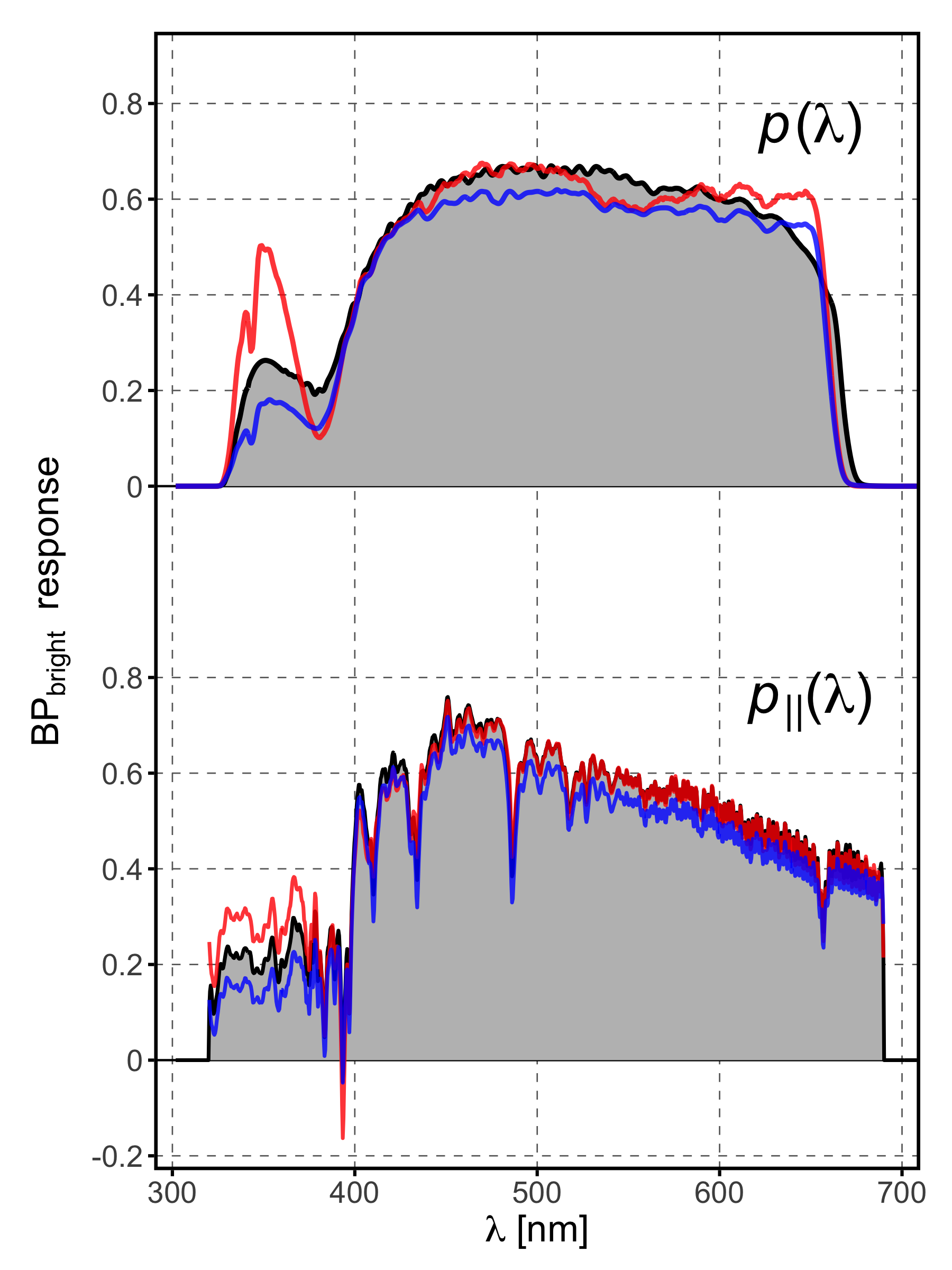}
   \includegraphics[width=0.48\textwidth]{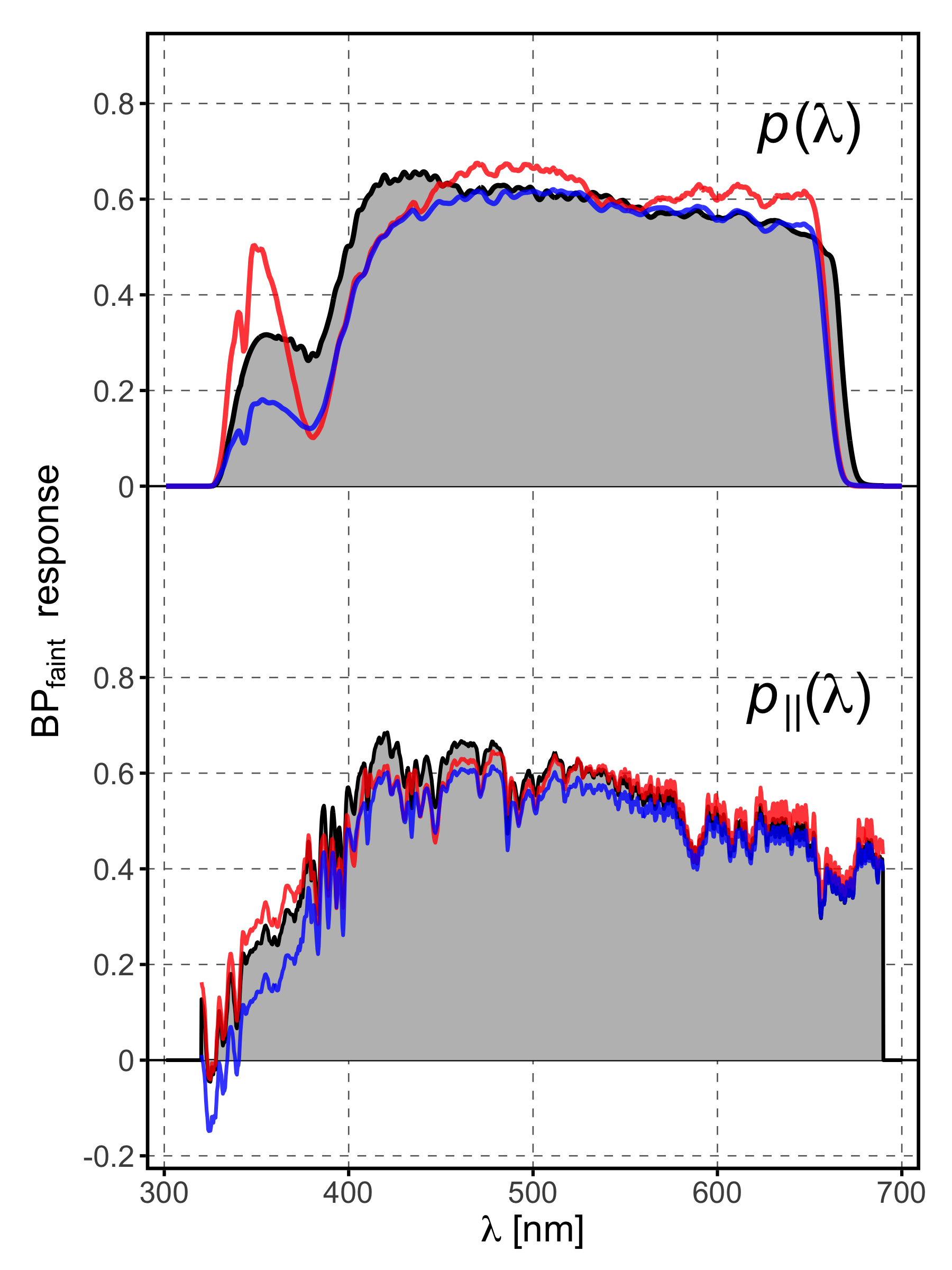}
   \caption{As Fig.~\ref{Fig:passbandG}, but for {\it BP}. Left: Solution for the bright magnitude regime. Right: Solution for the faint magnitude regime.}
              \label{Fig:passbandBP}
    \end{figure*}

For both the bright and the faint regime in $BP$ we use $M=4$ basis functions when computing the parallel component, and use cubic B-splines for the modification of the nominal pre-launch {\it BP} passband. The residuals obtained with the resulting passbands are also shown in Fig.~\ref{Fig:residualsBP}, right panel. A significant improvement can be achieved with the two passbands. The rms values of the residuals, included in Fig.~\ref{Fig:residualsBP}, improve for all four sets of spectra. Again, however, as the modifications introduced for the $BP$ passband correct for systematic effects in the calibration, the resulting differences depend on the particular sources considered, and the quantitative improvement cannot be generalised. The passbands for the bright and faint magnitude regimes are shown in Fig.~\ref{Fig:passbandBP}, together with the $BP_{REV}$ passband by \cite{Evans2018} and the {\it BP} passband by \cite{Jordi2010} for comparison. The basic parameters and zero points for the two {\it BP} passbands are listed in Table~\ref{tab:1}, the parallel and orthogonal components for both the bright and faint $BP$ passband are tabulated in Table~2.\par
The orthogonal components of the $BP$ passbands for the bright and faint regime were chosen in this work in such a way that the resulting passbands are similar to each other. Apart from the difference in shape, we find a difference of about 20~mmag in the zero points of the passbands for the two magnitude regimes.

\subsection{Comparison with other results}

Comparing the {\it BP} passbands derived in this work with previously published results, strong differences can be observed. Some of these differences occur only in the orthogonal component of the passband. Among these is the strong peak in the response of the $BP_{REV}$ passband around 350 nm. Such a strong, rather localised deviation between the initial expectation and the optimised passband is not required as far as the reproduction of the observed {\it BP} photometry is concerned. We therefore avoid such extreme peaks in the passband when estimating an orthogonal component, and give preference to less strong and, in wavelength space, less localised deviation from the initial guess for the passband. A second feature is the position of the wavelength cut-off position. This position is unconstrained by the calibration sources available in this work, and we obtained a solution that has a wavelength cut-off position slightly shifted towards longer wavelengths as compared to the passband by \cite{Evans2018} and \cite{Jordi2010}. As we cannot constrain the exact position of the cut-off wavelength from the available photometry, and, as discussed in the following section, the passbands derived in this work allow also for a good agreement in colour-colour relations beyond the space spanned by the calibration sources available for this work, we give preference to these solutions with somewhat larger cut-off wavelength.\par
Considering the parallel components of the $BP$ passbands shown in the lower panel of Fig.~\ref{Fig:passbandBP}, we find significant differences which are, in contrast to the differences introduced by the choice of the orthogonal component, relevant for obtaining a better reproduction of the observed photometry of the calibration sources. These differences mainly affect the short wavelength part of the two {\it BP} passbands. The differences with respect to the {\it REV} passbands are most likely caused by the separation of the sources into the bright and the faint magnitude regime. Deriving a single passband for the whole magnitude range includes systematic effects and results in a different and less suitable result in the parallel component as compared to this work.

\subsection{Colour-colour relations}

   \begin{figure*}
   \centering
   \includegraphics[width=0.98\textwidth]{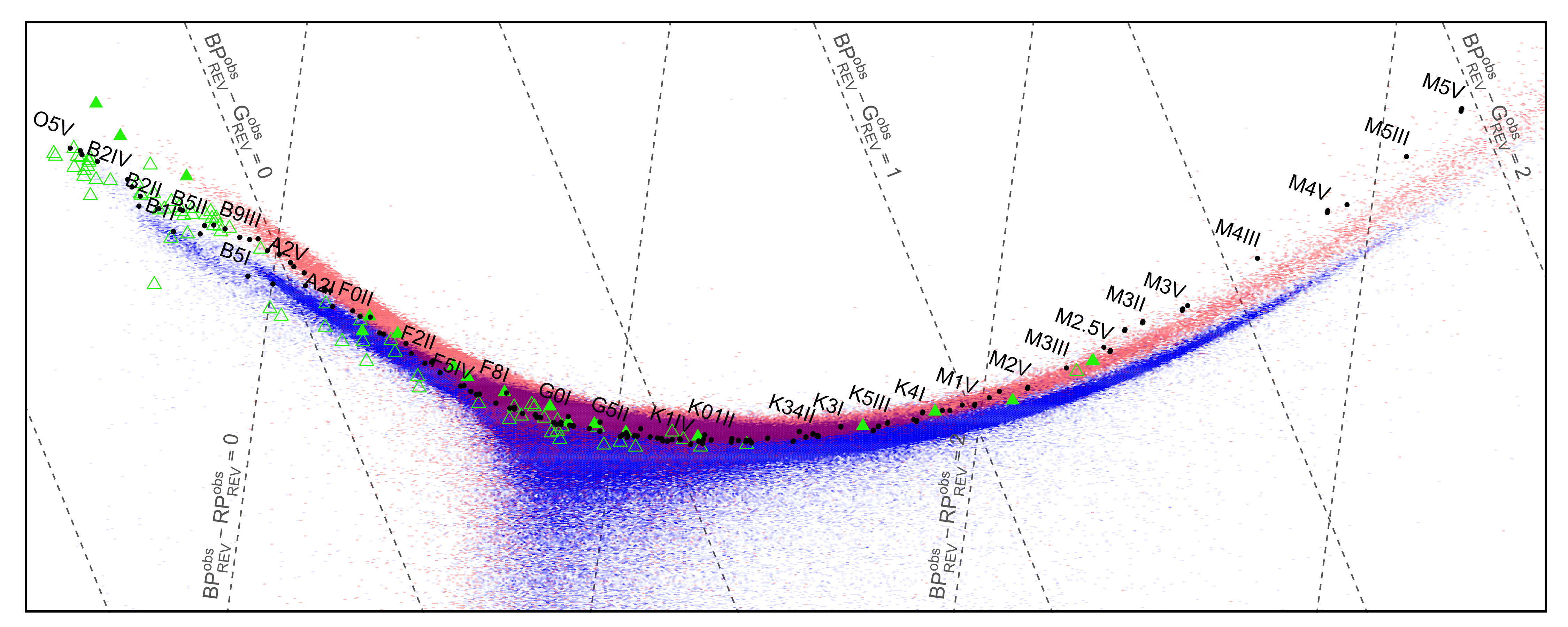}
   \includegraphics[width=0.98\textwidth]{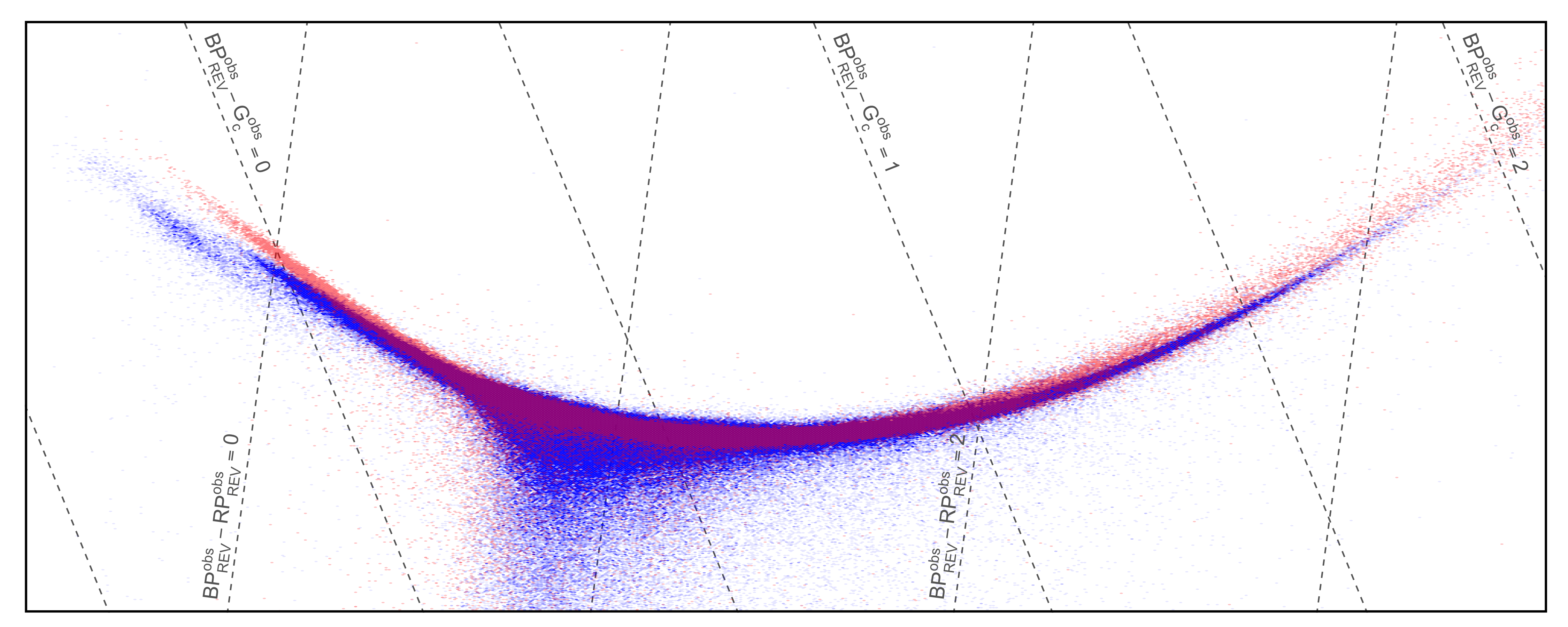}
   \includegraphics[width=0.98\textwidth]{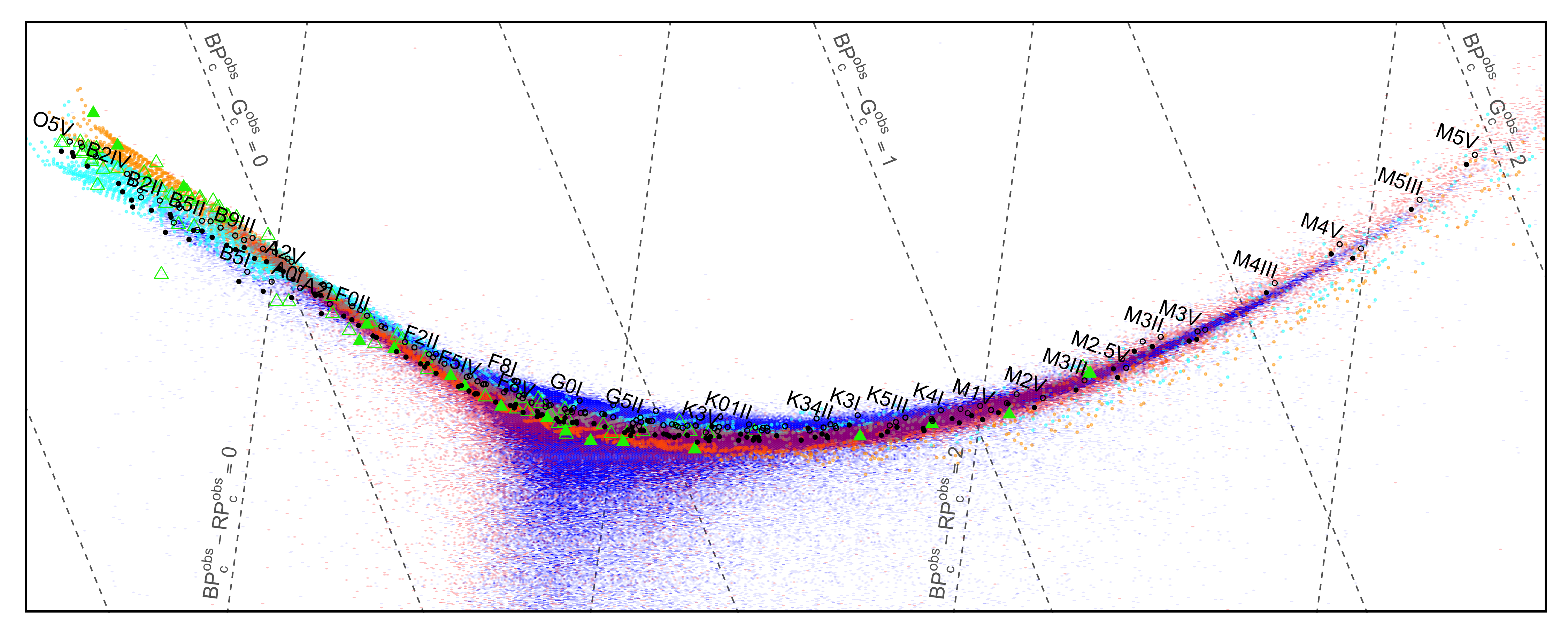}
   \caption{Colour-colour diagram for $BP-RP$ versus $BP-G$, separated into the bright (red shades) and faint (blue shades) magnitude range. Top panel: {\it REV} passbands. The black dots indicate the positions of the Pickles spectra, the green triangles show the SPSS (filled symbols for bright, open symbols for faint range). Central panel: As top panel, but replacing the $G_{REV}$ magnitude by the $G_c$ magnitude of this work. Bottom panel: Same as top panel, but for the $G_c$, $BP_c$, and $RP_c$ magnitudes of this work. The filled dots correspond to the Pickles spectra and the bright range {\it BP} passband, open dots for the {\it BP} passband for the faint magnitude range. The orange dots show the BaSeL spectra from the bright range, the cyan dots for the faint range. The original colour-colour diagram has been rotated by 30$^\circ$ clockwise and exaggerated in a vertical direction by a factor of 11 for better display. The dashed lines indicate the axis grid in $BP-RP$ and $BP-G$.} 
              \label{Fig:colourcolourblue}
    \end{figure*}

The split of the $BP$ photometric system into two regimes also becomes visible in colour-colour diagrams. Figure~\ref{Fig:colourcolourblue} shows the $BP - RP$ versus $BP - G$ diagram for the {\it Gaia} DR2 data set discussed in Sect.~\ref{sec:dr2}. For better fitting of the wide colour-colour diagram into the plotting window, it is rotated by 30$^\circ$ clockwise and the vertical axis is exaggerated by a factor of 11 as compared to the horizontal axis. The grid of dashed lines shows lines of constant values of $BP-RP$ and $BP-G$ for orientation. Sources from the bright and the faint magnitude regions are plotted in different colours for comparison. The top panel shows the data in the {\it REV} photometric system. The position of the SPSS and the \cite{Pickles} spectra are included in the plot for comparison. The most obvious difference in the distribution of the sources of the different magnitude regimes is an offset. This offset is mainly caused by the linear trend in $G$ described previously. Removing this trend reduces the differences between the colour-colour plots for the bright and faint sources strongly, as is shown in the central panel of Fig.~\ref{Fig:colourcolourblue}. In this panel, the $G_{REV}$ magnitude has been replaced by the $G_c$ magnitude. In the intermediate range of the diagram, the shape of the colour-colour relation of the bright and faint sources agrees very well. For very red sources, a slight difference between the relation of two magnitude regimes is present. The difference in trend for blue sources is obvious in this panel. Also the difference in zero point between the two passbands is visible. The convention in the VEGAMAG system requires that for all colour-colour diagrams, independently of the passbands, the main sequence runs through the point $(0,0)$. This requirement implies the need to shift in particular the faint sources curve upwards in Fig.~\ref{Fig:colourcolourblue} to meet the origin of the plot. The difference in shift for the bright and faint regime amounts to approximately 20~mmag. The difference between the bright and faint blue sources has already been pointed out by \cite{Arenou2018}, describing a larger dispersion for faint stars in the $BP - RP$ versus $BP - G$ diagram, and a jump of about 20~mmag around a $G \approx {\rm 11}$. \cite{Arenou2018} suspect the origin in the $G$ photometry but leave the reason for the observed effect unexplained. The results found in this work are thus in good agreement with previous findings. We attribute the larger dispersion for faint sources mainly to the linear drift in the $G$ photometry, and the jump of 20~mmag around $G \approx 11$ to the $BP$ photometry.\par
The bottom panel of Fig.~\ref{Fig:colourcolourblue} shows the colour-colour diagram for the $G_c$, $BP_c$, and $RP_c$ passbands derived in this work. This plot includes the correction for the zero points for the two $BP$ passbands. The distribution of the Pickles and BaSeL spectra in this diagram depends significantly on the choice of the $RP$ passband and is discussed in the following section.

\begin{table}
\begin{center}
\renewcommand\arraystretch{1.2}
\caption{\label{tab:1}Mean wavelength $\lambda_m$, pivot wavelength $\lambda_p$, zero points in the VEGAMAG and  AB photometric systems, and  $l_2$-norms of parallel and orthogonal components for  $G_c$, $BP_{c\,(bright)}$, $BP_{c\,(faint)}$, and $RP_c$ passbands derived in this work.}
\begin{tabular}{|l|c|c|c|c|} \hline
parameter & $G_c$ & $BP_{c\, (bright)}$ & $BP_{c\, (faint)}$ & $RP_c$ \\ \hline
$\lambda_m$ [nm] & 639.74 & 516.47 & 511.78 & 783.05 \\
$\lambda_p$ [nm] & 622.88 & 509.18 & 503.85 & 777.49 \\
zp [VEGA] & 25.6409 & 25.3423 & 25.3620 & 24.7600 \\
zp [AB] & 25.7455 & 25.3603 & 25.3888 & 25.1185 \\
$||p_\parallel||_2$ & 12.9612 & 9.7182 & 9.7241 & 11.5191 \\
$||p_\perp||_2$ & 2.0591 & 2.3606 & 2.1347 & 2.7351 \\ \hline
\end{tabular}
\end{center}
\end{table}

\section{\textit{RP} passband \label{sec:RP}}

\subsection{Determination of the passband}

\begin{figure*}
   \centering
   \includegraphics[width=0.48\textwidth]{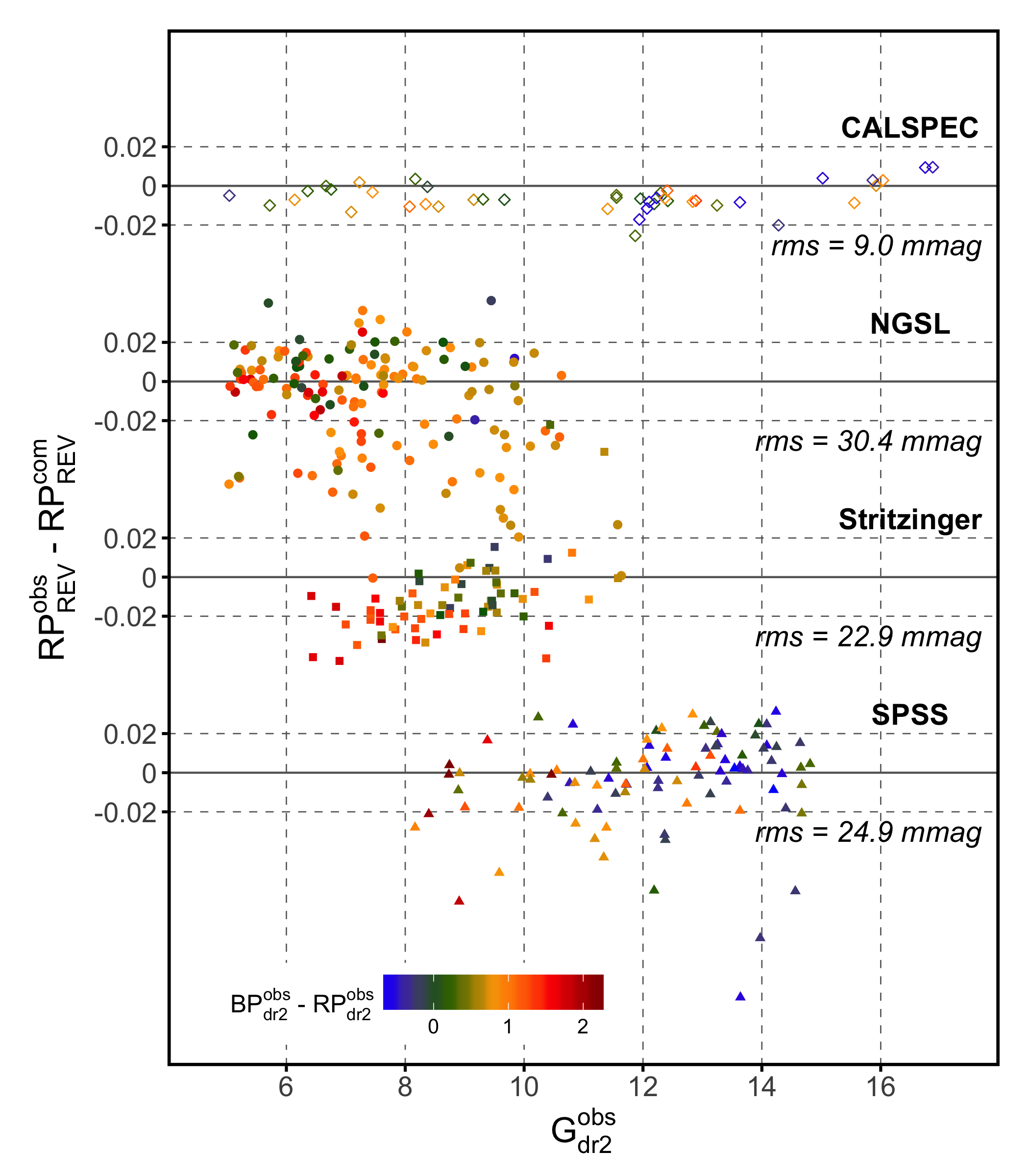}
   \includegraphics[width=0.48\textwidth]{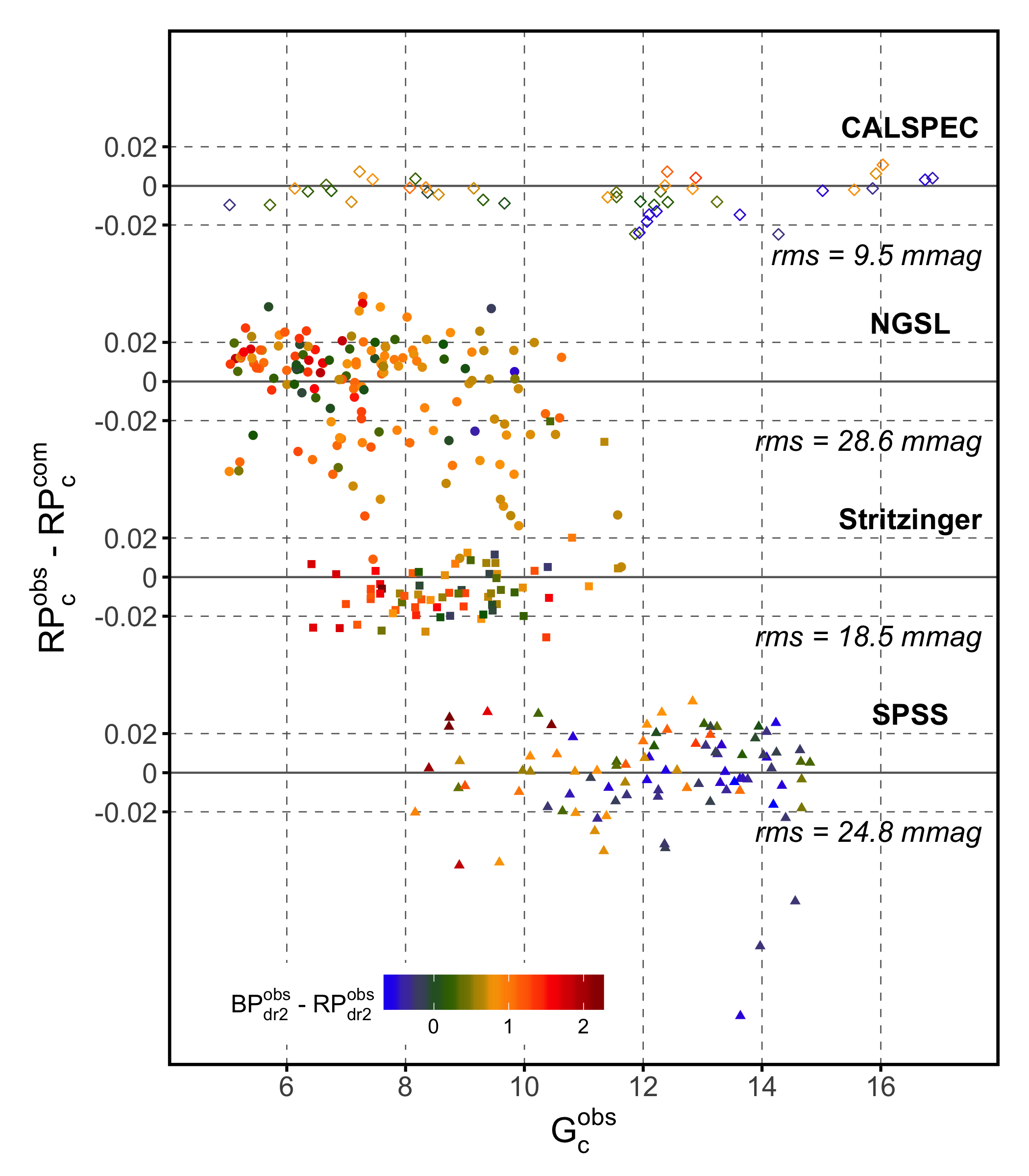}
   \caption{Residuals for the {\it RP} passband for four spectral libraries. Left panel: \cite{Evans2018}, {\it REV} passband. Right panel: This work. The rms of the residuals is presented for the four data sets.}
              \label{Fig:residualsRP}
    \end{figure*}

The residuals for the $RP_{REV}$ passband by \cite{Evans2018} are shown on the left panel of Fig.~\ref{Fig:residualsRP}. There is a slight tendency towards higher residuals for very red sources visible for the SPSS and Stritzinger data sets. The $RP_{dr2}$ and $RP_{REV}$ passbands were derived with the same set of SPSS calibration sources as used in this work. The modification of the initial $RP$ passbands by \cite{Evans2018} was done by multiplication with a first degree polynomial (see {\it Gaia} DR2 online documentation). This choice for a modification model might be too simplistic, not allowing us to produce a modified passband that has the optimal parallel component with respect to the calibration sources. As a consequence, a slight colour dependency of the residuals may have remained.\par
\begin{figure}
   \centering
   \includegraphics[width=0.48\textwidth]{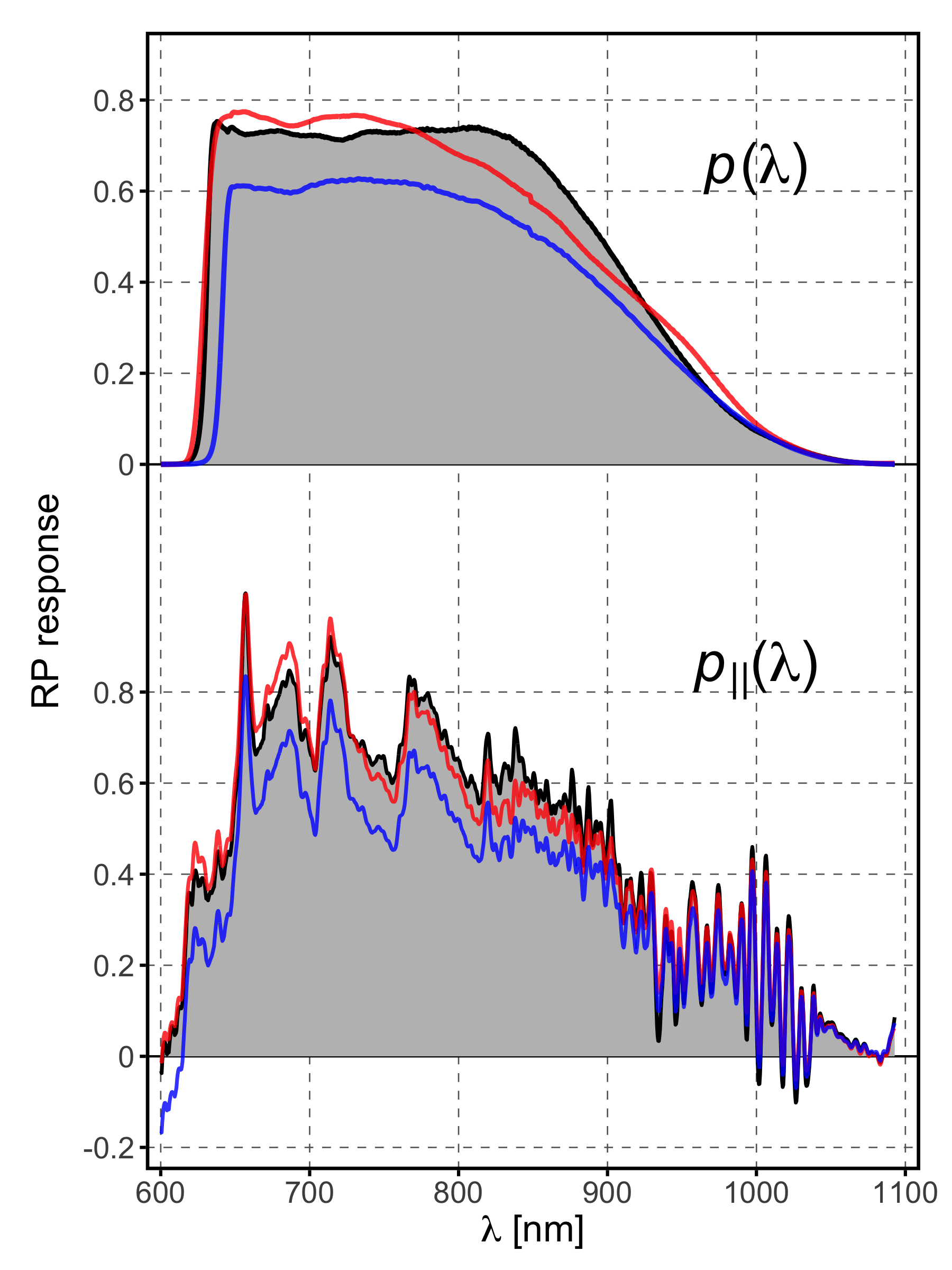}
   \caption{As Fig.~\ref{Fig:passbandG}, but for {\it RP}.}
              \label{Fig:passbandRP}
    \end{figure}

When computing the passband in this work, we cannot constrain the position of the wavelength cut-on position from photometry alone, as has already been the case for {\it BP} discussed before. As \cite{Evans2018} had spectroscopic observations from the {\it Gaia} {\it RP} instrument at hand before deriving the {\it RP} passband from the corresponding integrated photometry, the cut-on position of the passband could be constrained for the $RP_{REV}$ passband. We therefore chose an orthogonal component of the passband such that the wavelength cut-on position is the same as for the \cite{Evans2018} {\it REV} passband. We compute the parallel component of the $RP$ passband in this work using the SPSS calibration set with $M=5$ basis functions. The orthogonal component was estimated requiring the passband to reproduce the observational colour-colour relations, as described in Sect.~\ref{sec:colourcolourtheory} and discussed in more detail in Sect.~\ref{sec:RPcolourcolour}. The residuals for the solution of this work are shown in Fig.~\ref{Fig:residualsRP}, right panel. The passband is shown in Fig.~\ref{Fig:passbandRP}. The additional parameters for the passband are listed in Table~\ref{tab:1}, and the passband in Table 2. The passband solution derived in this work provides an improvement as compared to the {\it REV} passband, as it removes the systematic effect for very red sources in the residuals.

\subsection{Comparison with other results}

The {\it RP} passband derived in this work shows clear differences as compared to both the solution presented by \cite{Evans2018} and the pre-launch expectation. The differences affect both the parallel and orthogonal components of the passband. The {\it RP} passband derived in this work shows a steeper decrease at long wavelengths as compared to the other {\it RP} passbands presented before, and it is flatter at short and intermediate wavelengths. The parallel component derived in this work is lower than for the {\it REV} passband at short wavelengths, and larger at long wavelengths. The rms values of the residuals, listed in Fig.~\ref{Fig:residualsRP}, improve slightly for the SPSS, and more strongly for the NGSL and Stritzinger spectra. Only for the CALSPEC spectra, the $RP$ passband of this work results, for an unknown reason, in a slightly less good representation than the REV passband.

\subsection{Colour-colour relations \label{sec:RPcolourcolour}}

   \begin{figure*}
   \centering
   \includegraphics[width=0.49\textwidth]{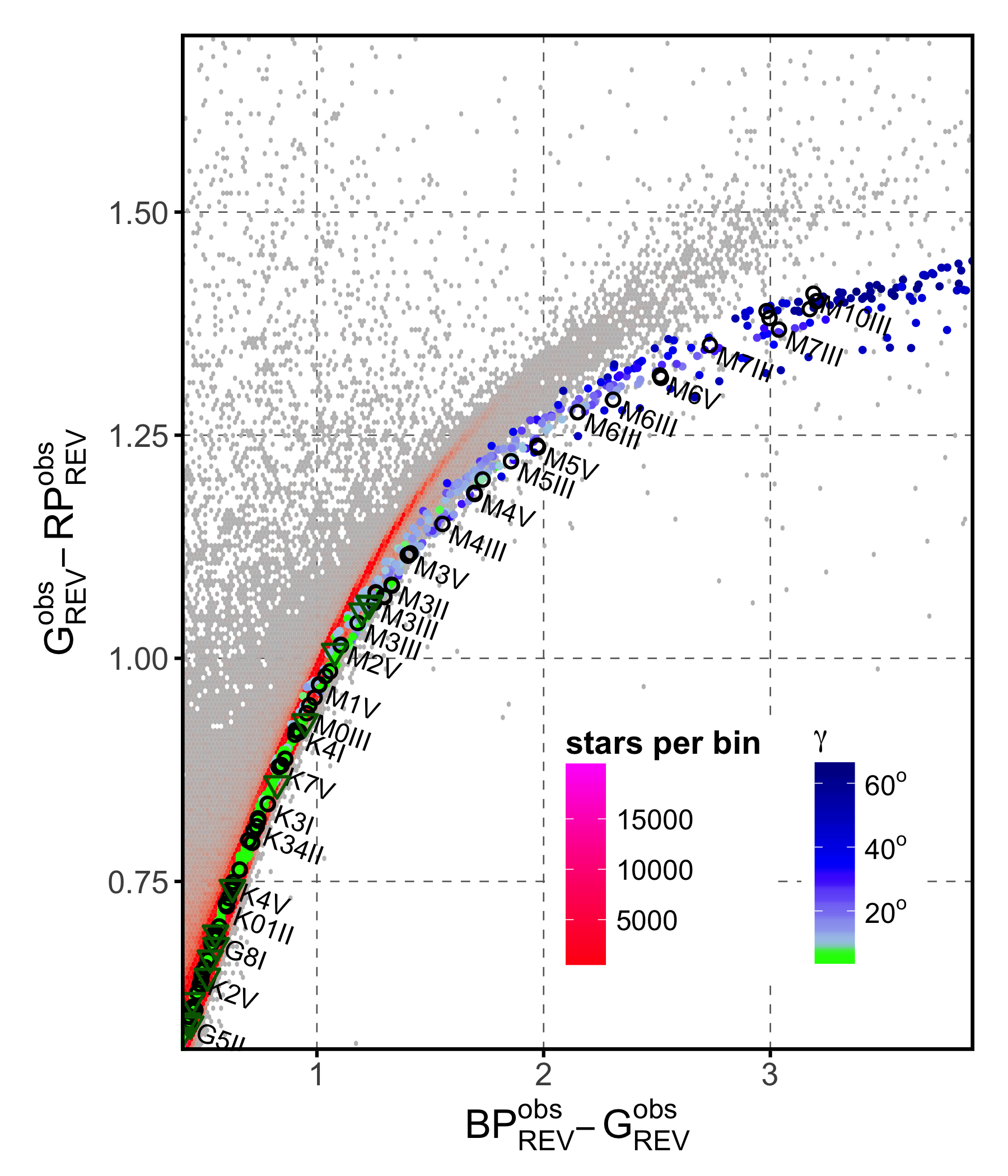}
   \includegraphics[width=0.49\textwidth]{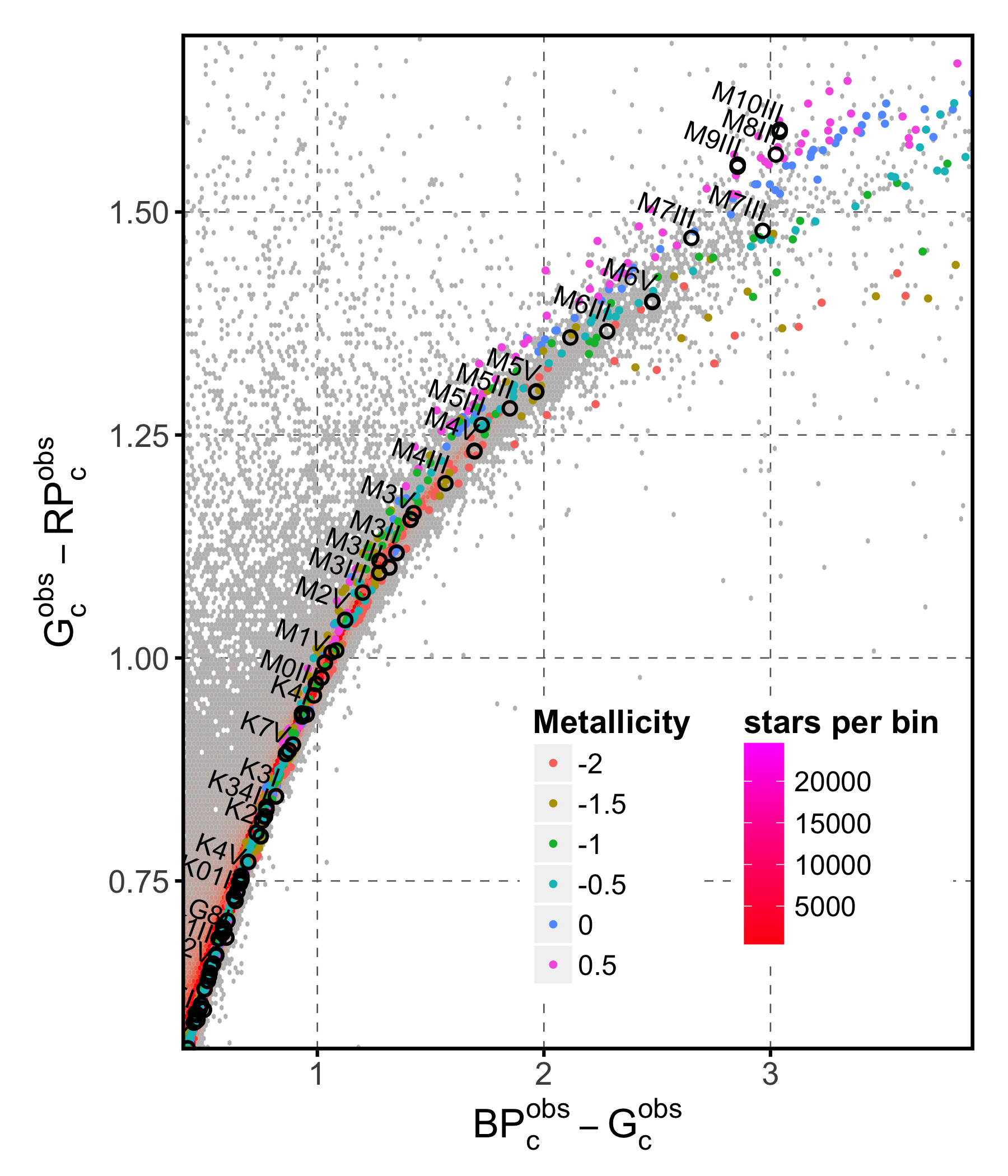}
   \caption{Left panel: Colour-colour diagram for the {\it REV} passbands. The black circles indicate the positions of the Pickles spectra, the dark green triangles show the SPSS. The dots represent the BaSeL spectra, colour-coded according to the estimated $\gamma$ angle. Right panel: Same as left panel, but for the $G_c$, $BP_c$, and $RP_c$ magnitudes of this work. The BaSeL spectra are shown for the faint magnitude range only, and are colour-coded according to the metallicity (in dex). Bins with numbers of stars lower than indicated in the colour bars are plotted in grey.}
              \label{Fig:colourcolourred}
    \end{figure*}

In Fig.~\ref{Fig:colourcolourblue} we compare the positions of the Pickles and BaSeL spectra in the $BP-RP$ versus $BP-G$ diagram. For the {\it REV} passbands, the synthetic follow approximately the observed relations in the colour-colour diagram, except for the very red sources. For a $BP-G$ colour index larger than approximately 1.25, a systematic deviation between the empirical and synthetic spectra and the {\it Gaia} DR2 observations occurs. This deviation starts for sources redder than the reddest SPSS available for calibration. This deviation can be explained by the influence of the orthogonal component of the passbands. M-type stars show strong variations in their SPDs, and very cool calibration sources are currently not available. As a consequence, the orthogonal component of the passband becomes increasingly important for the synthetic photometry of stars as their effective temperature becomes lower than the coolest calibration source. And if the guess for the orthogonal component is not optimal, this dependency on $p_\perp(\lambda)$ manifests itself in a progressive deviation of the synthetic colour-colour relations for Pickles and BaSeL spectra from the observed one. This effect may be quantified using the angle $\gamma$ as defined in \cite{Weiler2018}. For the Pickles and BaSeL spectra, however, the wavelength resolution of~500 is lower than the resolution of the SPSS spectra, which prevents an accurate computation of this quantity. For the BaSeL spectra, we introduced a colour coding according to $\gamma$ nevertheless, in order to illustrate the principle, in the $BP - G$ versus $G - RP$ colour-colour diagram in Fig.~\ref{Fig:colourcolourred}. The numerical values are to be considered as approximations only.\par
The red end of the $BP - G$ versus $G - RP$ colour-colour is shown in Fig.~\ref{Fig:colourcolourred}, using the {\it Gaia}~DR2 data set described in Sect.~\ref{sec:dr2}. The colour scale for the density of sources has been chosen to highlight the trend in the maximum of the distribution; all bins with lower star counts than indicated in the colour bar are presented in grey. This plot extends to even redder sources than Fig.~\ref{Fig:colourcolourblue}. In the left panel, showing the observed colour-colour relation and the comparison with Pickles and BaSeL spectra for the {\it REV} passbands, one can see the bifurcation of the distribution caused by the change in the $BP$ passband at $G_{dr2} \approx {\rm 10.99}$. As the very red SPSS belong to the bright magnitude range, the SPSS, Pickles, and BaSeL spectra follow the bright range distribution; the trend for the faint range distribution is seen on top of them. The increasing deviation between the observed colour-colour relation and the relation for the synthetic spectra is also obvious, showing a systematic behaviour down to the coolest Pickles sources of spectral type M10. For the BaSeL spectra, the colour scale indicates the estimated $\gamma$ angle. As three different angles are of relevance in the colour-colour diagram, which are the ones with respect to the $G$, $BP$, and $RP$ passbands, the largest value of the three has been used in the colour scale for each BaSeL SPD shown. For BaSeL SPDs with colour indices up to the redest SPSS, the agreements between the synthetic and observed colour-colour relations are good, and $\gamma$ remains low. For SPDs redder than the reddest SPSS, the systematic deviation starts, and $\gamma$, indicating the dependence on the orthogonal components of the three passbands involved, increases progressively as the deviation increases. This illustrates the use of $\gamma$ for estimating the sensitivity of a spectrum with respect to the choice of $p_\perp(\lambda)$.\par
The right panel in Fig.~\ref{Fig:colourcolourred} shows the same data set for the passbands derived in this work. The bifurcation is removed in this diagram by the use of the $G_c$ and $BP_c$ passbands. As no calibration sources cooler than M2V, the coolest SPSS in the calibration data set, are available, we constrain the $RP$ passband by requiring it not only to result in a good agreement between observed and synthetic colour-colour relations for the calibration sources, but also to result in an agreement between observed colour-colour relations and the relations for the Pickles spectra. The $RP$ passband derived in this work achieved such an agreement for sources up to a spectral type of M7. For the Pickles spectra of types M8 to M10, a slight deviation from the observed relationships remains. The synthetic colour-colour relation of the BaSeL spectra also shows a better agreement with the observational relation. The scatter between the different BaSeL spectra, however, is very large for very red sources. This scatter mainly results from the differences in metallicity between the spectra, as is indicated by the colour coding of the BaSeL points in the right hand panel of Fig.~\ref{Fig:colourcolourred}. It remains uncertain if this spread according to metallicity of cool M-type sources is a real effect or whether it results from an inaccurate estimate of the orthogonal passband components. It is possible to derive passband solutions that reduce the spread among the BaSeL spectra, but only at the cost of either a worse agreement between the BaSeL and Pickles spectra in the colour-colour diagram, or by introducing a more complex shape in the $RP$ passband. We chose in this work passband solutions that result in a simple shape of the $RP$ passband and a better agreement between Pickles and BaSeL spectra, rather than reducing the spread according to metallicity for late M-type stars. This again illustrates the need for very cool calibration sources in order to allow for a reliable interpretation of the photometry of M-type stars.


\section{Summary and conclusions \label{sec:summary}}

In this work, we determined the {\it G}, {\it BP}, and {\it RP} passbands for {\it Gaia} DR2. We used a functional analytic formulation of the problem of passband reconstruction to separate each passband into a sum of two functions. One of these functions, the parallel component $p_\parallel(\lambda),$ is fully constrained by the calibration sources, while the other function, the orthogonal component $p_\perp(\lambda),$ is fully unconstrained. We derived the parallel component using sets of observational spectral libraries and the {\it Gaia} DR2 photometry of the sources included in the libraries. The orthogonal component is estimated based on two considerations. First, we use an initial guess for the passband, based on a-priori knowledge on the passband shape. This initial guess is then modified in such a way that the modified initial guess remains close to the initial guess while having $p_\parallel(\lambda)$ in agreement with the determination. Second, we introduce an additional constraint on the choice of the orthogonal component by requiring the passbands to predict colour-colour relationships for model spectra of a wide range of spectral types to be in agreement with the observed colour-colour distributions in {\it Gaia} DR2.\par
For the $G$ passband, we found a solution for the shape of the passband that is basically in agreement with the shape of solutions published by \cite{Evans2018} and \cite{Jordi2010}. An approximately linear trend in the $G$ photometry is however present on a magnitude interval from about 6 to 17. A trend in $G$ photometry has already been observed in {\it Gaia} Data Release~1 \citep{Weiler2018}, and it was already noticed for {\it Gaia} DR2 by \cite{Evans2018} and \cite{Arenou2018}. The origin of this effect remains yet unknown. In this work, we introduce a correction for it by applying a linear correction in $G$ magnitude of $\rm 3.5$~mmag per magnitude. This correction results in a factor of 0.9965 in the computation of the $G_c$ magnitude according to Eq. (\ref{eq:Gmag}). This correction applies to a magnitude interval from about 6 to 17 in $G$ band. For brighter sources, a deviation from this trend caused by saturation effects occurs, as reported by \cite{Evans2018}. For magnitudes larger than 17, no suitable calibration sources are available for this work. \cite{Arenou2018}, however, found indications for a more complex systematics in the residuals for sources fainter than about 17.\par
For $BP$ we introduced two different photometric systems for stars brighter and fainter than 10.99 in the $G_{dr2}$-band. By doing so, we take into account an inconsistency in the $BP$ photometry, which occurs for a reason as yet unknown, but it may be connected to a change in instrumental configuration such as gate activation or windowing. The assumption of two $BP$ passbands explains the differences between the $G$ and $BP$ photometry previously reported by \cite{Arenou2018}. We derive two passbands for $BP$, valid for sources brighter than 10.99 mag in $G_{dr2}$, and fainter than this limit. The $BP$ passbands derived for the bright and the faint magnitude range are rather similar to each other in shape, and differ in zero point by about 20~mmag from each other. However, they differ significantly from previously published passbands by \cite{Evans2018} and \cite{Jordi2010}.\par
The passband for $RP$ provides an improvement as compared to the passband by \cite{Evans2018}, as it removes a systematic colour-dependent effect in reproducing the calibration sources. The passband is rather different in shape from the previously published $RP$ passbands.\par
The passband solutions for $G$, $BP$, and $RP$ in this work have been chosen such that, when applying them together, the synthetic photometry from the \cite{Pickles} spectral library is in good agreement with the colour-colour relations observed in {\it Gaia}~DR2. The passbands presented in this work result in synthetic colour-colour relations that follow well the observed relations for stars from spectral type O5 to M7. For cooler sources than M7, the database becomes too sparse and the variation within synthetic colour-colour relations too large to draw definite conclusions. The different passbands derived in this work are thus consistent with each other within the limits set by the available calibration data.\par
We point out that the shapes of the passbands derived in this work are not unique, as is the case for the passbands derived by \cite{Evans2018}. The limitation is introduced by the existence of the orthogonal component and is of a fundamental nature. There necessarily exist an infinite number of different passbands that describe all available calibration data equally well. In order to estimate the sensitivity of a particular SPD with respect to the orthogonal component, we suggest the computation of the contribution of $p_\parallel(\lambda)$ and $p_\perp(\lambda)$ to the synthetic photometry independently. For this purpose, all passbands derived in this work are presented in Table 2 with their parallel and orthogonal components individually. The angle $\gamma$ as defined in Eq.~\ref{eq:gamma} may serve as an illustrative quantity for specifying the degree to which a particular SPD depends on the orthogonal component and, with it, how sensitive the synthetic photometry of the given SPD is to systematic errors that may arise from an inaccurate estimate of $p_\perp(\lambda)$.\par
The passbands presented in this work allow for a better agreement between synthetic and observed photometry over a wide range of SPDs than previously published passbands, while at the same time being physically reasonable. Table~2 lists all passbands derived in this work as a function of wavelength and is available in electronic form at the Centre de Donn{\'e}es astronomiques de Strasbourg, CDS.  Applying them in the interpretation of the {\it Gaia} DR2 photometric data may therefore help to maximise the scientific outcome from the uniquely large and accurate astronomical data set that  is {\it Gaia} DR2.

\begin{acknowledgements}
      This work was supported by the MINECO (Spanish Ministry of Economy) through grants ESP2016-80079-C2-1-R (MINECO/FEDER, UE) and ESP2014-55996-C2-1-R (MINECO/FEDER, UE) and MDM-2014-0369 of ICCUB (Unidad de Excelencia ''Mar{\'i}a de Maeztu'').\\
      This work has made use of data from the European Space Agency (ESA) mission {\it Gaia} (\url{https://www.cosmos.esa.int/gaia}), processed by the {\it Gaia} Data Processing and Analysis Consortium (DPAC, \url{https://www.cosmos.esa.int/web/gaia/dpac/consortium}). Funding for the DPAC has been provided by national institutions, in particular the institutions participating in the {\it Gaia} Multilateral Agreement.\\
      I thank the {\it Gaia} DPAC SPSS team for providing the intermediate data products used for this work: E. Pancino, G. Altavilla, S. Marinoni, N. Sanna, G. Cocozza, S. Ragaini, and S. Galleti.\\
      I furthermore thank C. Jordi, J.~M. Carrasco and C. Fabricius for fruitful discussions during the preparation of this work. 
\end{acknowledgements}

%
   \bibliographystyle{aa} 
   \bibliography{DR2PassbandsRef} 
%

\end{document}